\documentclass[reprint,showpacs,preprintnumbers,amssymb,prb,aps]{revtex4-1}

\usepackage{epsf}
\usepackage{epsfig}
\usepackage{graphicx}
\usepackage{subfigure}
\usepackage{amsmath}
\usepackage{braket} 
\usepackage{enumerate}

\DeclareMathOperator*{\tr}{Tr}
\DeclareMathAlphabet\mathbfcal{OMS}{cmsy}{b}{n}

\begin{document}

\title{Interaction-induced charge and spin pumping
through a quantum dot at finite bias}

\author{Hern\'{a}n L. Calvo}

\affiliation{Institut f\"{u}r Theorie der Statistischen Physik, RWTH
Aachen University, 52056 Aachen, Germany}
\affiliation{JARA - Fundamentals of Future Information Technology}

\author{Laura Classen}

\affiliation{Institut f\"{u}r Theorie der Statistischen Physik, RWTH
Aachen University, 52056 Aachen, Germany}
\affiliation{JARA - Fundamentals of Future Information Technology}

\author{Janine Splettstoesser}

\affiliation{Institut f\"{u}r Theorie der Statistischen Physik, RWTH
Aachen University, 52056 Aachen, Germany}
\affiliation{JARA - Fundamentals of Future Information Technology}

\author{Maarten R. Wegewijs}

\affiliation{Institut f\"{u}r Theorie der Statistischen Physik, RWTH
Aachen University, 52056 Aachen, Germany}
\affiliation{Peter Gr\"{u}nberg Institut, Forschungszentrum J\"{u}lich,
52425 J\"{u}lich, Germany}
\affiliation{JARA - Fundamentals of Future Information Technology}

\begin{abstract}
We investigate charge and spin transport through an adiabatically
 driven, {\em strongly interacting} quantum dot weakly coupled to two
 metallic contacts with {\em finite bias voltage}. Within a kinetic
 equation approach, we identify coefficients of response to the
 time-dependent external driving and relate these to the concepts
 of charge and spin emissivities previously discussed within the
 time-dependent scattering matrix approach. Expressed in terms of
 auxiliary vector fields, the response coefficients allow for a
 straightforward analysis of recently predicted interaction-induced
 pumping under periodic modulation of the gate and bias voltage
 [Phys. Rev. Lett. {\bf 104}, 226803 (2010)]. We perform a detailed
 study of this effect and the related adiabatic Coulomb blockade
 spectroscopy, and, in particular, extend it to spin pumping. Analytic
 formulas for the pumped charge and spin in the regimes of small and
 large driving amplitude are provided for arbitrary bias. In the absence
 of a magnetic field, we obtain a striking, simple relation between the
 pumped charge at zero bias and at bias equal to the Coulomb charging
 energy. At finite magnetic field, there is a possibility to have
 interaction-induced {\em pure spin pumping} at this finite bias value,
 and generally, additional features appear in the pumped charge. For
 large-amplitude adiabatic driving, the magnitude of both the pumped
 charge and spin at the various resonances saturate at values which are
 independent of the specific shape of the pumping cycle. Each of these
 values provide an independent, quantitative measurement of the junction
 asymmetry.
\end{abstract}


\maketitle

\section{Introduction}

The generation of a dc current through a mesoscopic system is usually
associated to a bias voltage maintained between the contacts to the
exterior world. Remarkably, charge and spin transport can even be
achieved in the absence of an external bias by the cyclic
modulation of some of the parameters of the system.\cite{Thouless83}
When this modulation is slow compared to the characteristic dwell time
of the electrons, the transport mechanism is called {\em adiabatic
pumping}. Here, the pumped charge is of geometric nature, since it depends
on the specific shape of the path sustained by the system's parameters
but not on its detailed time
evolution.\cite{Altshuler99,Avron00,Makhlin01,Zhou03} For appropriate
modulation setups,\cite{Niu84, Avron01} the pumped charge after one
period may be quantized in units of the electron
charge,\cite{Kouwenhoven91,Pothier92,Chorley12} motivating
its use as a highly precise current standard for quantum
metrology\cite{Keller98} or in the initialization \cite{Feve07} and
processing of coherent states \cite{Beenakker05, Samuelsson05,
Splettstoesser09, Sherkunov12} in the context of quantum information.
In the opposite limit where the transferred charge is not necessarily
quantized, the pumping mechanism is dominated by quantum
interference\cite{Zhou99} of the coherent electrons in the
device. In the last years, adiabatic pumping was widely studied both
experimentally \cite{Switkes99, Watson03, Fletcher03} and
theoretically. In noninteracting systems, a well-established theory was
formulated by Brouwer.~\cite{Brouwer98} It makes use of the concept of
emissivity which was introduced in the scattering matrix approach for
time-dependent systems at low frequency by B\"uttiker, Thomas and
Pr\^{e}tre.\cite{Buttiker94} Importantly, this formalism is
adequate as long as interactions can be described on a self-consistent
mean-field level.\cite{Buttiker93} Within this formalism, several
aspects of adiabatic pumping were explored, covering diverse effects
such as dissipation and noise,\cite{Makhlin01,Moskalets02} or spin
polarized pumping.\cite{Mucciolo02} Further works dealt with different
setups including normal metal-superconducting
heterostructures,\cite{Wang01} pumping by surface acoustic
waves,\cite{Levinson00} and graphene-based quantum pumps.\cite{Prada09} 

Pumping through confined electron systems dominated by a strong Coulomb
interaction is a particularly challenging topic since the mean-field
approach breaks down and a new formulation is necessary. Several studies
addressed interaction effects in specific setups and 
regimes.\cite{Citro03, Aono04, Schiller08, Arrachea08,
Aleiner98, Brouwer05, Cota05, Splettstoesser05, Sela06,
Splettstoesser06, Splettstoesser08, Cavaliere09, Winkler09,
Reckermann10, Riwar10, Reckermann12, Kashuba12, Riwar12} In
Ref.~\onlinecite{Citro03}, pumping is investigated in interacting
quantum wires. By using a slave boson mean field approximation, Aono
studied adiabatic pumping through a quantum dot in the Kondo
regime.\cite{Aono04} This regime was also treated in the Toulouse
limit\cite{Schiller08} and for nonadiabatic 
pumps\cite{Arrachea08} by using the Keldysh Green's function
technique. Pumping through open quantum dots was described by employing
bosonization techniques.\cite{Aleiner98,Brouwer05} In the Coulomb
blockade regime, spin pumping was addressed through a numerical
calculation of the reduced density matrix of a double quantum
dot.\cite{Cota05} An expression for the adiabatic pumping current in
interacting systems was derived using a nonequilibrium Green's function
technique in Refs.~\onlinecite{Splettstoesser05, Sela06}. A diagrammatic
real-time approach,\cite{Splettstoesser06} was used to investigate
several aspects of adiabatic pumping through
weakly-coupled interacting quantum-dot systems\cite{Splettstoesser08,
Cavaliere09, Winkler09, Reckermann10, Riwar10, Riwar12, Reckermann12}
and served as the basis for non-equilibrium renormalization group
studies that treat the tunneling non-perturbatively.\cite{Kashuba12}

Among the above mentioned studies, only a few discuss the modulation of
the applied bias.\cite{Moskalets04} In particular, \emph{pumping around a
nonequilibrium working point} induced by a static nonlinear bias was
addressed in noninteracting systems.\cite{Entin-Wohlman02}
Recently, a strongly interacting single-level quantum dot with a
modulation of the gate and bias voltage was
investigated.\cite{Reckermann10} For this modulation setup, on top of a
dc current produced by the bias, an additional adiabatic dc current is 
generated by the Coulomb interaction. Interestingly, this
interaction-induced pumping current can be accessed by using lock-in
techniques, and was suggested as a new spectroscopic tool to probe
internal properties of the system, like spin degeneracy and junction
asymmetries. Similar effects were reported\cite{Yuge12} for an open
quantum system when controlling the temperatures and chemical potentials
of the reservoirs. In Ref.~\onlinecite{Riwar12}, the zero-frequency
pumping noise in adiabatically driven quantum dots is discussed for
time-dependent bias, revealing further information on the tunnel
coupling asymmetry in cases where the pumped charge is zero.

In this paper, we investigate interaction-induced charge pumping in
detail and extend it to the spin degree of freedom. We focus on the
interplay between the strong local Coulomb interaction in a quantum dot
and the nonequilibrium effects induced by finite bias and the modulation
around this working point. To describe the dynamics of the local system,
the coupling to the leads and the frequency of the modulation are
treated perturbatively.\cite{Splettstoesser06} In particular, we 
restrict ourselves to the single-electron tunneling (SET) regime.

We derive a general expression for the adiabatic charge and spin
currents in response to a change in the internal occupations of the dot
induced by the driving parameters. The coefficients of this response of
the current are related to the emissivities to the leads,\cite{Buttiker94}
establishing a connection to Brouwer's pumping theory \cite{Brouwer98}
which is applicable as long as the single-particle picture
holds. Moreover, we write the pumped charge and spin as the flux
generated by auxiliary vector fields in the space of the
parameters. This allows us to find the general conditions under which a
finite pumped charge or spin may occur, independent of details specific
to the model or the pumping cycle. 

We apply this strategy for the case of a modulation of the gate and bias
voltage and analyze the influence of the local interaction on the
generation of pumping. The recently introduced ``stability diagram'' for
the pumped charge\cite{Reckermann10} is shown to be readily understood in
terms of these vector fields and extended to pumped spin. A detailed
analysis of these diagrams is provided for all the regimes of the
applied bias, including analytic fitting formulas for the pumped 
charge and spin in the limits of weak and large driving amplitude. We
find that, although not quantized, the pumped charge (and spin) for
large driving amplitudes saturates at plateau values which depend on the
asymmetry in the coupling to the leads and the working region of the
applied voltages. At zero magnetic field, a surprising and simple
relation between the pumped charge at high-bias and at zero-bias is
found. It is exploited for a quantitative determination of the junction
asymmetry by two single measurements. In a finite magnetic field, the
spin and charge pumping are not anymore trivially connected at
high-bias. In this regime, pure spin pumping occurs when the coupling to
the leads is symmetric.

Our paper is organized as follows. In the next section we introduce the
model and review the theoretical framework used in the calculation of
the pumped charge and spin. In Sec.~\ref{sec:results}, we consider the
pumped charge and spin for the specific modulation of the gate and bias
voltage, first concentrating on the role of the interaction and then, in
Sec.~\ref{sec:Bfinite}, we discuss the effects induced by an external
magnetic field. We summarize our results in Sec.~\ref{sec:summary}.

\section{Model and Formalism}

\subsection{Model}

\begin{figure}[tbp]
\includegraphics[width=0.4\textwidth]{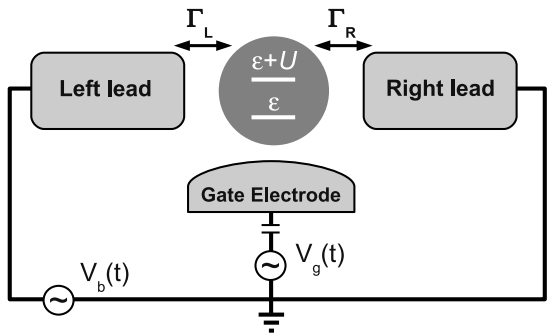}
\caption{Scheme of the considered model. The interacting quantum dot
 (grey circle) is capacitively coupled to two leads via $\Gamma_L$ and
$\Gamma_R$. The transport is controlled through time-dependent gate
 and bias voltages. The lines in the dot indicate the transition
 energies $\epsilon_N-\epsilon_{N-1}$, $N=1,2$, for zero external
 magnetic field $B$.}
\label{fig:scheme}
\end{figure}

We consider a single-level quantum dot with Coulomb interaction weakly
coupled to two noninteracting leads as sketched in
Fig.~\ref{fig:scheme}. The full system, containing the dot, the left
($L$) and right ($R$) leads, and the tunneling between dot and leads, is
described by the total Hamiltonian $H(t) =
H_\mathrm{dot}(t)+H_\mathrm{res}(t)+H_\mathrm{tun}$. The quantum dot
Hamiltonian is given by
\begin{equation}
H_\mathrm{dot}(t) = \sum_\sigma \epsilon_\sigma(t) \hat{n}_\sigma + U
 \hat{n}_\uparrow \hat{n}_\downarrow, 
\end{equation}
where we denote the spin-resolved number operator by $\hat{n}_\sigma =
d_\sigma^\dag d_\sigma$ and $U$ is the Coulomb charging energy. The
fermionic operator $d_\sigma^\dag$ ($d_\sigma$) creates (annihilates) an
electron in the dot with spin $\sigma = \uparrow,\downarrow$. The
many-body eigenstates of the dot are characterized by their charge
number and spin by $\ket{0}$ for an empty dot,
$\ket{\sigma}=d_\sigma^\dag \ket{0}$ for a singly occupied dot with spin
$\sigma$ and $\ket{2}=d_\downarrow^\dag d_\uparrow^\dag \ket{0}$ for a
doubly occupied dot. Their energies are then $0,\epsilon_\sigma$ and
$\epsilon_\uparrow + \epsilon_\downarrow + U$, respectively, where
$\epsilon_\sigma(t) = \epsilon(t)-\sigma B/2$.
The level position $\epsilon(t)=-\alpha V_g(t)$ is capacitively
modulated by the time-dependent gate voltage $V_g(t)$, with lever arm
$\alpha < 1$. Furthermore, $B$ accounts for the Zeeman splitting
produced by an external magnetic field $\mathbf{B} = B \mathbf{e}_z$ in
units $e = \hbar = k_B = g \mu_B = 1$. Here we use $\sigma = \pm 1$ as a
convenient notation for spin $\uparrow$ and $\downarrow$, respectively.
The underlying capacitive description of the parameters $U$ and $\alpha$
is sketched in Fig.~\ref{fig:circuit} and will be discussed below. 

The leads are described as reservoirs of noninteracting electrons
through the Hamiltonian
\begin{equation}
H_\mathrm{res}(t) = \sum_{rk\sigma} [\epsilon_{rk}+\mu_r(t)]
 c_{rk\sigma}^\dag c_{rk\sigma},
\end{equation}
where $c_{rk\sigma}^{\dag}$ ($c_{rk\sigma}$) creates (annihilates) an
electron in the lead $r = L,R$ with spin $\sigma = \uparrow,\downarrow$
and state index $k$. The eigenenergies of the leads are uniformly
shifted by the time-dependent bias voltage $V_b(t)$ such that the
electro-chemical potentials read $\mu_r(t) = \pm V_b(t)/2$ for $r =
L,R$. These reservoirs are furthermore characterized by a temperature $T$.

Finally, the tunnel coupling between the dot and the leads is determined
by the tunnel Hamiltonian
\begin{equation}
H_\mathrm{tun} = \sum_{rk\sigma} V_r d_\sigma^\dag
 c_{rk\sigma}+h.c.,
\end{equation}
with the tunnel matrix element $V_r$, which we assume to be independent
of $k$ and $\sigma$. The tunnel-coupling strength $\Gamma_r = 2\pi
|V_r|^2 \nu_r$ characterizes the rate at which tunnel processes take
place. Here $\nu_r$ is the density of states in the $r$-lead, which is
assumed to be energy-independent and with a band cutoff $D_r$, which is
the largest energy scale.

We are interested in the simultaneous modulation of gate and bias
voltages. These are driven around a time-independent working point
specified by $\bar{V}_g$ and $\bar{V}_b$ with a fixed relative phase
$\Delta\phi = \phi_g - \phi_b$:
\begin{equation}
V_x(t) = \bar{V}_x+\delta V_x \sin(\Omega t+\phi_x), \:
 x = g,b,
\label{eq:voltages}
\end{equation}
where $\delta V_x$ is the driving amplitude. At time $t_0$, we ``switch
on'' the coupling between the dot and the leads and calculate the
time-dependent nonequilibrium steady state at a much later time
$t$. This includes two sources of nonequilibrium: the finite bias and
the adiabatic driving. On top of the stationary current flow, an
additional time-dependent current, which can have a dc component, is
generated by the periodic modulation. We work in the adiabatic regime
where the driving period $\mathcal{T} = 2\pi/\Omega$ is larger than the
typical time spent by the electrons inside the quantum dot. Here, the
modulation frequency and the driving amplitude are limited by the {\em
adiabaticity condition} $\alpha \delta V_g, \delta V_b \ll T^2 /\Omega$.

\subsubsection{Observables} 

In the description of the dot occupancies through the reduced density
matrix, we assume the reservoirs to be always in equilibrium. Since
leads and dot are decoupled at the initialization time $t_0$, the total
density matrix is factorized in both subsystems as $\hat{\rho} =
\hat{p}_\mathrm{res} \hat{p}$, with the density matrix of the leads
given by the one of a grand canonical ensemble
\begin{subequations}
\begin{align}
 \hat{p}_\mathrm{res} &= \prod_r \frac{1}{z_r} e^{- \left[H_r(t') -
 \mu_r(t')\hat{N}_r\right]/T},\\ 
 z_r &= \tr_r e^{- \left[H_r(t')-\mu_r(t')\hat{N}_r \right]/T},
\end{align}
\end{subequations}
and $t' \leq t_0$. Here $\hat{N}_r=\sum_{k\sigma} c_{rk\sigma}^\dag
c_{rk\sigma}$ is the particle-number operator in the $r$-lead and the
temperature $T$ is assumed to be the same in the two leads. Notice that
the density operator of the reservoirs remains time-independent, meaning
that the occupation in the leads is not affected by the modulation of
the bias voltage. For electric field modulations with
a frequency well below the plasma frequency of the leads
(tens of $\mathrm{THz}$ in typical doped semiconductors) this assumption
is well justified.\cite{Jauho94} 

The time evolution of the expectation value of an arbitrary operator
$\hat{R}$ is formally obtained by
\begin{equation}
 R(t) = \braket{\hat{R}}(t) = \tr_{\mathrm{dot}} \tr_{\mathrm{res}}
  \left(\hat{R} \, \hat{\rho}(t) \right). 
\label{eq:currents}
\end{equation}
In the next sections, we will describe the tunneling currents $I_r(t) =
\braket{\hat{I}_r}(t)$ and $J_r(t) = \braket{\hat{J}_r}(t)$ related to
the charge and the spin component along the external field $B$,
respectively, entering the $r$-lead. Since in the uncoupled system $H_0
= H_\mathrm{dot}+H_\mathrm{res}$ the number of particles and the
$z$-component of spin are conserved, the operators related to these
observables are given by  
\begin{subequations}
\begin{align}
\hat{I}_r &= i[H_\mathrm{tun},\hat{N}_r],\\
\hat{J}_r &= i[H_\mathrm{tun},\hat{S}_z^r],
\end{align}
\end{subequations}
with $\hat{S}_z^r = \sum_{k\sigma} \frac{\sigma}{2} c_{rk\sigma}^\dag
c_{rk\sigma}$. Since we consider a system that preserves rotation
symmetry around the magnetic field ($z$-) axis, the $x$ and $y$
components of the spin vector may only occur as a transient effect and
are not required here.

\subsection{Real-time diagrammatic approach}

In this section we outline the theoretical framework used to identify
the adiabatic contribution to the time-resolved currents of
Eq.~({\ref{eq:currents}}). As we will show in Eqs.~(\ref{eq:adcurrent})
and (\ref{eq:rescoeff}), the adiabatic current can be interpreted as the
delayed response of the dot occupation probabilities to the driving of
the external parameters. The full relevance of this result will become clear
in the next section. Eq.~(\ref{eq:rescoeff}) allows to apply the concept of {\em
emissivity} to a system with strong Coulomb interaction and arbitrary
bias. Originally,\cite{Buttiker94} the emissivity was introduced to
describe capacitive effects on time-dependent response on a
self-consistent mean-field level and was further used on adiabatic
pumping through non- or weakly interacting systems.~\cite{Brouwer98,Sela06}

We start with the description of the relevant part of the dot's reduced
density matrix, namely, its diagonal elements, obtained after tracing
out the degrees of freedom of the leads. The time evolution
of the dot occupation probabilities, represented by the vector
$\mathbf{p}(t) =
(p_0(t),p_\uparrow(t),p_\downarrow(t),p_2(t))^\mathrm{T}$, is governed
by the generalized master equation\cite{Splettstoesser06}  
\begin{equation}
 \frac{d}{dt}\mathbf{p}(t) = \int_{-\infty}^t
 dt'\mathbf{W}(t,t')\mathbf{p}(t').
 \label{eq:kineq}
\end{equation}
Here, the change in the dot occupation probabilities, due to electron
tunnel processes between dot and leads, is accounted by the kernel
$\mathbf{W}(t,t')$. In terms of the real-time diagrammatic technique
developed in Ref.~\onlinecite{Konig96}, this kernel collects all
irreducible diagrams in the Keldysh double contour. Its matrix elements
$W_{m,n}(t,t')$ describe the transition from a state $\ket{n}$ at time
$t'$ to a state $\ket{m}$ at time $t$. It is important to notice that
the transport properties are completely determined by the diagonal
elements of the reduced density operator. In the chosen basis the
off-diagonal elements, related to coherent superposition of different
states, are decoupled from diagonal ones due to charge and spin
conservation in the tunneling and therefore do not affect the currents.

Since we consider a weakly coupled system bounded to a slow modulation
of the energy levels, it is sufficient to describe the lowest order
contribution in the tunnel coupling and in the time-dependent
perturbation introduced by the driving. The occupation probabilities are
thus expanded in powers of $\Omega$ by $\mathbf{p}(t) =
\mathbf{p}_t^{(i)}+\mathbf{p}_t^{(a)}$, bearing in mind that
$\Omega/\Gamma \ll 1$. The first term (zeroth-order in $\Omega$)
represents the {\em instantaneous} occupations and describes the steady
state solution when the parameters are frozen at time $t$. Here the
index $t$ indicates the parametric time-dependence through the driving
parameters $\{\chi(t)\}$, i.e. $\mathbf{p}_t^{(i)} =
\mathbf{p}^{(i)}(\{\chi(t)\})$. The instantaneous occupations are
obtained from the time-dependent kinetic equation in the stationary limit
\begin{equation}
\mathbf{0} = \mathbf{W}_t^{(i)} \mathbf{p}_t^{(i)},
\label{eq:kineqin}
\end{equation}
together with the normalization condition $\mathbf{e}^\mathrm{T}
\mathbf{p}_t^{(i)} = 1$, where $\mathbf{e} = (1,1,1,1)^\mathrm{T}$,
and we introduced the zero-frequency Laplace transform of the
instantaneous kernel $\mathbf{W}_t^{(i)} = \int_{-\infty}^t dt'
\mathbf{W}^{(i)}(t-t')$. In the SET regime we consider here,
characterized by the linear dependence on $\Gamma$ of
$\mathbf{W}_t^{(i)}$, the result coincides with the one from Fermi's
golden rule. The next-to-leading term (linear in $\Omega$),
$\mathbf{p}_t^{(a)}$, obeys the {\em adiabatic} correction to the
kinetic equation
\begin{equation}
 \frac{d}{dt} \mathbf{p}_t^{(i)} = \mathbf{W}_t^{(i)}
 \mathbf{p}_t^{(a)}.
\label{eq:kineqad} 
\end{equation}
The retardation correction $\mathbf{p}_t^{(a)}$ is determined by the
competition between the driving (left hand side) and the inverse
response times contained in $\mathbf{W}_t^{(i)}$ (right hand side). The dot
occupation probabilities are obtained by solving Eqs.~(\ref{eq:kineqin})
and (\ref{eq:kineqad}) together with the normalization condition
$\mathbf{e}^\mathrm{T}\mathbf{p}_t^{(i)} = 1$ and
$\mathbf{e}^\mathrm{T}\mathbf{p}_t^{(a)} = 0$. From
Eq.~(\ref{eq:kineqad}), the adiabatic corrections to the occupation
probabilities are written in terms of the instantaneous contributions by 
\begin{equation}
\mathbf{p}_t^{(a)} = \left[\tilde{\mathbf{W}}_t^{(i)}\right]^{-1}
 \frac{d}{dt} \mathbf{p}_t^{(i)}, 
\label{eq:occ_ad}
\end{equation}
where the (invertible) matrix 
\begin{equation}
\left[\tilde{\mathbf{W}}_t^{(i)}\right]_{ij} =
 \left[\mathbf{W}_t^{(i)}\right]_{ij} -
 \left[\mathbf{W}_t^{(i)}\right]_{ii}, 
\end{equation}
includes the normalization condition $\mathbf{e}^\mathrm{T}
\mathbf{p}_t^{(a)} = 0$.

The charge and spin currents in Eq.~(\ref{eq:currents}) need to be
equally expanded in both the frequency $\Omega$ and the tunnel-coupling
strength $\Gamma$. The resulting observables are then split into
instantaneous and adiabatic correction terms
\begin{equation}
  R_t^{(i/a)} = \braket{\hat{R}}_t^{(i/a)} = \mathbf{e}^\mathrm{T}
  \mathbf{W}_{R,t}^{(i)} \mathbf{p}_t^{(i/a)},
\label{eq:obs}
\end{equation}
where $\mathbf{W}_{R,t}^{(i)}$ is the instantaneous kernel of the
corresponding current $R$ which, in the present approximation, is linear 
in $\Gamma$. We describe $R_t^{(a)}$ by a scalar product with the
time-derivative of the dot state occupation probabilities 
\begin{subequations}
\begin{align}
R_t^{(a)} &= \mathbf{e}^\mathrm{T} \mathbf{W}_{R,t}^{(i)}
 \left[\tilde{\mathbf{W}}_t^{(i)}\right]^{-1} \frac{d}{dt}
 \mathbf{p}_t^{(i)} \\ 
&:=\left[\boldsymbol{\varphi}_t^R \right]^\mathrm{T} \frac{d}{dt}
 \mathbf{p}_t^{(i)} \\
&= \sum_j \varphi_{j,t}^R \frac{d}{dt}p_{j,t}^{(i)},
\end{align}
\label{eq:adcurrent}%
\end{subequations}
with the sum running over the dot eigenstates,
i.e. $j=0,\uparrow,\downarrow,2$. Applied to the adiabatic charge and 
spin currents, this equation defines the adiabatic current $R$ as the
response to a time-dependent variation in the instantaneous occupation
probabilities induced by the external modulation. The response
coefficient, 
\begin{equation}
\varphi_{j,t}^{R} = \frac{\partial R_t^{(a)}}{\partial
 \dot{p}_{j,t}^{(i)}},
\label{eq:rescoeff}%
\end{equation}
determines the ratio at which the current $R$ flows into the $r$-lead
due to a variation in the occupation of the state $j$.
The relevance of these coefficients lies in the fact that
they distinguish the amount of charge (or spin) that enters into each one
of the leads. As compared to the instantaneous solution, the response
coefficients give information about the characteristic delay time for
the current $R$.\cite{Kashuba12}

\subsection{Adiabatically pumped charge and spin}

Now that we have an explicit expression for the adiabatic current, we
can determine the charge and spin pumped through the quantum dot during
one modulation cycle. The purpose of this section is to relate the
resulting pumped charge and spin to the emissivity of the
contacts,\cite{Buttiker94} a well-known concept from scattering
theory. It measures the amount of charge entering the $r$-lead due to 
the variation $\delta\chi$ of the driving parameter $\chi$. Following
the reasoning by B\"uttiker {\em et al},\cite{Buttiker94} for a slow
variation of $\chi$, the charge entering the $r$-lead is related to the
emissivity $dN(r)/d\chi$ by
\begin{equation}
\delta Q_{I_r} = \frac{dN(r)}{d\chi} \delta\chi.
\label{eq:deltaq}
\end{equation}
We are interested in the adiabatic\cite{note1} pumped charge when
varying two parameters $\chi_1(t)$ and $\chi_2(t)$ over a cycle of the
driving. Calculated as the integral of Eq.~(\ref{eq:deltaq}), this
reads\cite{Brouwer98} 
\begin{equation}
 Q_{I_r} = \int_0^\mathcal{T} dt \left(
 \frac{dN(r)}{d\chi_1} \frac{d\chi_1}{dt} +
 \frac{dN(r)}{d\chi_2}
 \frac{d\chi_2}{dt} \right). 
\label{eq:brouwer} 
\end{equation}
In terms of the adiabatic current of Eq.~(\ref{eq:adcurrent}), this last
can also be written as
\begin{subequations}
\begin{align}
 Q_{I_r} &= \int_0^\mathcal{T} dt I_{r,t}^{(a)}\\
 &= \sum_j \int_{0}^\mathcal{T} dt \,
 \varphi_{j,t}^{I_r} \frac{d}{dt} p_{j,t}^{(i)}. 
\end{align}
\label{eq:pumping}%
\end{subequations}
Since $p_{j,t}^{(i)}$ depends on $t$ through the driving parameters, we
rewrite its time-derivative in terms of $\dot{\chi}_{1,2}$. A comparison
with Eq.~(\ref{eq:brouwer}) allows one to relate the emissivity with the 
response coefficients and the occupation probabilities by
\begin{equation}
 \frac{dN(r)}{d\chi} = \sum_j \varphi_j^{I_r}
 \frac{\partial p_j^{(i)}}{\partial \chi},
\label{eq:emc}
\end{equation}
where $\chi$ is either $\chi_1$ or $\chi_2$ and the index $t$ is removed
to emphasize that $\varphi_j^{I_r}$ and $p_j^{(i)}$ are functions of
$\chi$ rather than $t$. Analogously, the above relation can also be
generalized for the {\em spin} emissivity\cite{Brataas12} $dS(r)/d\chi$
in terms of the spin current response coefficients as follows 
\begin{equation}
 \frac{dS(r)}{d\chi} = \sum_j \varphi_j^{J_r}
 \frac{\partial p_j^{(i)}}{\partial \chi}.
\label{eq:ems}
\end{equation}
Written in this way, the emissivity is the weighted rate of change in
the occupation probabilities due to the external perturbation. The
response coefficients $\varphi_j^{I_r}$ and $\varphi_j^{J_r}$ describe
the rate at which the charge and the spin, respectively, are transferred
to the leads when the occupation probabilities are changed by the driving
parameters. The above Eqs.~(\ref{eq:emc}) and (\ref{eq:ems}), also shown
by Sela and Oreg\cite{Sela06} for adiabatic transport at equilibrium,
extend the known result from scattering matrix
theory~\cite{Buttiker94,Brouwer98} to a system with strong Coulomb
interaction driven around a nonequilibrium steady state.

\subsubsection{Pseudo vector potential and pseudo magnetic field}

We now describe the pumped charge and spin in terms of auxiliary vector
fields defined in the space of the driving parameters. Our purpose here
is to relate these vector fields with the response coefficients of
Eq.~(\ref{eq:rescoeff}). As we will show in the next section, using
these fields we can conveniently describe the conditions for finite
pumping and provide a detailed insight into the ``stability diagrams''
for the pumped charge and spin.

According to Eqs.~(\ref{eq:brouwer}) and (\ref{eq:pumping}), the charge
and spin pumped in a cycle of the modulation can be written as the line
integral 
\begin{equation}
Q_R = \oint_{C} d\boldsymbol\chi \cdot
 \mathbfcal{A}_R(\boldsymbol\chi).
\end{equation}
For the two-dimensional parameter space studied here, spanned by
$\mathbf{e}_{\chi_1} = (1,0)$ and $\mathbf{e}_{\chi_2} = (0,1)$, the
position in the closed trajectory $C$ is indicated by
$\boldsymbol\chi = \sum_i \chi_i \mathbf{e}_i$. This integral is
independent of how fast the path is traversed, and consequently the
total pumped charge (or spin) does not depend on the driving
frequency as long as the adiabaticity condition is fulfilled.
The geometric aspects of the problem, entering through the field
$\mathbfcal{A}_R$, are certainly of
interest.~\cite{Avron00,Zhou03,Cohen03} However, here they are merely
convenient auxiliary quantities to analyze the problem of
interaction-induced pumping. In analogy to classical electrodynamics, 
the vector field
\begin{equation}
 \mathbfcal{A}_R(\boldsymbol\chi) = \sum_j \varphi_j^R(\boldsymbol\chi)
 \boldsymbol{\nabla} p_j^{(i)}(\boldsymbol\chi), 
\label{eq:pseudoa}
\end{equation}
with $\boldsymbol{\nabla} = \sum_i \partial_{\chi_i} \mathbf{e}_i$, can be
interpreted as a {\em pseudo} vector potential defined in the space of
the driving parameters. From Eqs.~(\ref{eq:emc}) and (\ref{eq:ems}), the
components of this vector potential are given by the emissivities to the
leads, i.e. $\mathbfcal{A}_{I_r} = dN(r)/d\boldsymbol\chi$ and
$\mathbfcal{A}_{J_r} = dS(r)/d\boldsymbol\chi$. Therefore, the vector
potentials describe, respectively, the amount of charge and spin
entering the leads due to the change in the driving
parameters. From the form of $\mathbfcal{A}_R$ in Eq.~(\ref{eq:pseudoa}),
we notice that for constant response coefficients, the resulting pseudo
vector potential is just a gauge function $\mathbfcal{A}_R = \sum_j
\boldsymbol{\nabla} (\varphi_j^R p_j^{(i)})$ which, integrated over a closed
trajectory, gives zero pumping.

Using Stokes' theorem we write the pumped charge and spin in terms of
the surface integral
\begin{equation}
 Q_R = \iint_\Sigma d\mathbf{S} \cdot \mathbfcal{B}_R(\boldsymbol\chi),
 \label{eq:QR}
\end{equation}
where $\Sigma$ is any area in the parameter space encircled by $C$, such
that $C = \partial\Sigma$. Written in this way, we can imagine the
pumped charge and spin as the {\em flux} generated by a {\em pseudo}
magnetic field
\begin{subequations}
\begin{align}
\mathbfcal{B}_R(\boldsymbol\chi) &= \boldsymbol{\nabla} \times \mathbfcal{A}_R
 (\boldsymbol\chi)\\ 
 &= \sum_j \boldsymbol{\nabla} \varphi_j^{R}(\boldsymbol\chi) \times
 \boldsymbol{\nabla} p_j^{(i)}(\boldsymbol\chi).
\end{align}
\label{eq:pseudob}%
\end{subequations}
The advantage of this representation is that the pseudo magnetic field
{\em anticipates} the conditions for finite pumping without referring
to the specific details of the model and the modulation. The direction
of this field is, by construction, perpendicular to the plane
(i.e. pointing outside the parameter space) defined by the driving parameters,
i.e. $\mathbfcal{B}_R (\boldsymbol\chi) = \mathcal{B}_R(\boldsymbol\chi)
\, \mathbf{e}_{\chi_1} \times \mathbf{e}_{\chi_2}$. For a fixed
direction of the driving, the sign of the pumped charge is given by the
sign of $\mathcal{B}_R$, which depends on the internal details of the
system (e.g., Coulomb interaction, coupling to the leads, spin
degeneracy). We stress that this interpretation of $Q_R$ comes
purely from the adiabaticity condition in the time-dependent parameters,
and should not be confused with any other effect due to the external
magnetic field $B$ eventually present in this setup. 

\section{Results}
\label{sec:results}

In this section, we apply the above theory for the specific case of a
single-level quantum dot slowly driven by the gate and bias voltage. In
particular, we want to understand how the pumped charge and spin are
affected by the interplay between the local Coulomb interaction and
nonequilibrium effects induced by a modulation around a finite bias.
To this end, we introduce expressions for the adiabatic charge and spin
currents for a general modulation and then we take the specific
modulation of the voltages given in Eq.~(\ref{eq:voltages}).

Our starting point is the description of the time-resolved adiabatic
charge current given in Eq.~(\ref{eq:adcurrent}). This is obtained from
the explicit calculation of the matrix elements of the kernels
$\mathbf{W}_t^{(i)}$ and $\mathbf{W}_{I_r,t}^{(i)}$ introduced in the
previous section. In the evaluation of the response coefficients related
to the occupation of the different states of the dot [see
Eq.~(\ref{eq:rescoeff})], we find out that it is possible to combine
them in such a way that the adiabatic current can be written in terms of
the time-derivatives of the instantaneous average charge
$\braket{\hat{n}}_t^{(i)} = p_{\uparrow,t}^{(i)} +
p_{\downarrow,t}^{(i)} + 2p_{2,t}^{(i)}$ and spin
$\braket{\hat{S}_z}_t^{(i)} =
(p_{\uparrow,t}^{(i)}-p_{\downarrow,t}^{(i)})/2$. We recover the
result~\cite{Reckermann10}
\begin{equation}
I_{r,t}^{(a)} = \varphi_{n,t}^{I_r}
 \frac{d}{dt} \braket{\hat{n}}_t^{(i)} +
 \varphi_{S_z,t}^{I_r} \frac{d}{dt}
 \braket{\hat{S}_z}_t^{(i)},
 \label{eq:ccurnt_ad}
\end{equation}
where the charge current response coefficients, related to the charge
and spin in the dot, respectively, read
\begin{subequations}
 \begin{align}
  \varphi_{n,t}^{I_r} &= -\frac{(\Gamma-\gamma)(\Gamma_r+\gamma_r) +
  \beta\beta_r}{\Gamma^2 - \gamma^2+\beta^2},\\
  \varphi_{S_z,t}^{I_r} &= -2\frac{(\Gamma_r + \gamma_r)\beta -
  (\Gamma+\gamma)\beta_r}{\Gamma^2-\gamma^2+\beta^2}.
 \end{align}
 \label{eq:crescoeff}%
\end{subequations}
together with $\gamma = \sum_r \gamma_r$, $\beta = \sum_r \beta_r$, and
$\Gamma = \sum_r \Gamma_r$. The response coefficients depend
parametrically on $t$ through the following factors
\begin{subequations}
 \begin{align}
  \gamma_r &= \frac{\Gamma_r}{2} \sum_\sigma \left[
  f\left(\frac{\epsilon_{r\sigma}}{T}\right) -
  f\left(\frac{\epsilon_{r\sigma}+U}{T}\right)\right],\\
  \beta_r &= \frac{\Gamma_r}{2} \sum_\sigma \sigma \left[
  f\left(\frac{\epsilon_{r\sigma}}{T}\right) -
  f\left(\frac{\epsilon_{r\sigma}+U}{T}\right)\right]. 
 \end{align}
 \label{eq:rates}%
\end{subequations}
Since the adiabatic charge currents flowing into each one of the leads
are related by particle conservation, i.e. $\sum_r I_{r,t}^{(a)} =
-\frac{d}{dt}\braket{\hat{n}}_t^{(i)}$, no net charge is accumulated on
the dot after one period of the modulation. 

Noticeably, the above expression for the adiabatic charge current is not
restricted to a particular choice of the driving parameters. In this
sense, Eq.~(\ref{eq:ccurnt_ad}) is valid for arbitrary combinations
involving not only the voltages $V_g(t)$ and $V_b(t)$ but also
$\Gamma_r(t)$, $B(t)$ (in a fixed direction), $U(t)$, etc., because the
eigenstates of the system remain time-independent. For the specific
modulation we consider here, given by Eq.~(\ref{eq:voltages}), the time
dependence exclusively enters in the arguments 
\begin{equation}
\epsilon_{r\sigma}(t) = \epsilon_\sigma(t)-\mu_r(t),
\end{equation}
of the Fermi function $f(\omega) = [1+\exp(\omega)]^{-1}$. We now extend
the above calculation for the adiabatic spin current. From the
evaluation of the matrix elements of $\mathbf{W}_{J_r,t}^{(i)}$ we
obtain
\begin{equation}
J_{r,t}^{(a)} = \varphi_{n,t}^{J_r}
 \frac{d}{dt} \braket{\hat{n}}_t^{(i)} +
 \varphi_{S_z,t}^{J_r} \frac{d}{dt}
 \braket{\hat{S}_z}_t^{(i)}.
 \label{eq:scurnt_ad}
\end{equation}
In this case, the spin current response coefficients are
\begin{subequations}
\begin{align}
\varphi_{n,t}^{J_r} &= -\frac{1}{2} \frac{(\Gamma-\gamma)\beta_r -
 (\Gamma_r-\gamma_r)\beta}{\Gamma^2-\gamma^2+\beta^2},\\
\varphi_{S_z,t}^{J_r} &=
 -\frac{(\Gamma+\gamma)(\Gamma_r-\gamma_r)+\beta_r\beta}
 {\Gamma^2-\gamma^2+\beta^2}. 
\end{align}
\label{eq:srescoeff}%
\end{subequations}
For the spin-isotropic quantum dot discussed here, the adiabatic spin
current only occurs in the presence of an external magnetic field. For
$B = 0$ at any time, rotation symmetry implies $\braket{S_z}_t^{(i)} =
0$ and $\beta_r 
= 0$, such that the current vanishes for all times. The spin currents
$J_{L,t}^{(a)}$ and $J_{R,t}^{(a)}$ are related by spin conservation,
i.e. $\sum_r J_{r,t}^{(a)} = -\frac{d}{dt}\braket{\hat{S_z}}_t^{(i)}$,
meaning that no accumulation of spin is allowed after one cycle of the
parameter modulation. 

\begin{figure}[tbp]
 \includegraphics[width=0.4\textwidth]{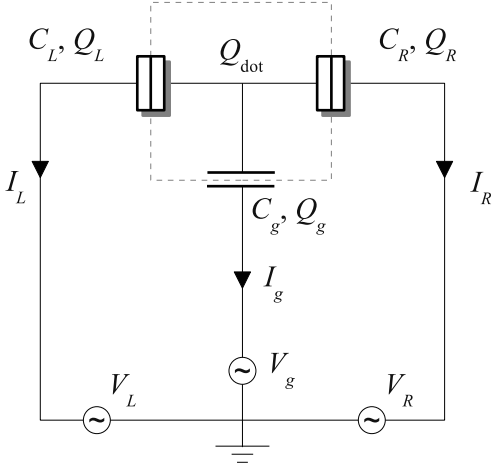}
 \caption{Schematics of the equivalent circuit for the proposed
 model. $C_L, C_R$ are the capacitances between the dot and the leads
 and $C_g$ is the capacitance between the dot and the gate electrode. The
 quantum dot region is indicated by dashed lines.}
 \label{fig:circuit}
\end{figure}

\subsection{Displacement current and pseudo gauge invariance}

In addition to the tunneling currents we introduced above, a
displacement current:
\begin{equation}
 I_r^\mathrm{dis}(t) = \frac{d}{dt} Q_r^\mathrm{scr}(t),
\end{equation}
generally occurs in the present setup due to the moving screening charges
$Q_r^\mathrm{scr}$ in the gate and the leads. These arise in response to
a variation in the electrostatic potential induced by a change of the
charge on the dot. To describe them, we consider the Coulomb-blockade
model,\cite{Averin91} in which the system is represented by the
equivalent circuit shown in Fig.~\ref{fig:circuit}. The screening
charges are given by the difference
\begin{equation}
 Q_r^\mathrm{scr}(t) = C_r \left[ V_r(t) -
 V_\mathrm{dot}(Q_\mathrm{dot}(t)) \right],
\end{equation}
between the applied voltage at the $r$-lead and the electrostatic
potential inside the dot (in Fig.~\ref{fig:circuit}, the region
delimited by dashed lines). $C_r$ is the capacitance between the dot and
the $r$-lead. Due to particle conservation we have
$\dot{Q}_\mathrm{dot}(t) = d\braket{\hat{n}}/dt = -[I_L(t) + I_R(t)]$
and the displacement current reads\cite{Bruder94}
\begin{equation}
 I_r^\mathrm{dis}(t) = C_r \dot{V}_r(t) - \frac{C_r}{C} \sum_{r'} \left[
 C_{r'} \dot{V}_{r'}(t) + I_{r'}(t) \right],
\label{eq:dispcrnt}
\end{equation}
where $C = C_L + C_R + C_g$ is the total capacitance. As in the case of
the tunneling currents, we can separate the displacement current into
instantaneous and adiabatic contributions. Since the time-derivatives of
the applied voltages are of linear order in $\Omega$, they do not enter in
the instantaneous term and 
\begin{equation}
 I_{r,t}^{\mathrm{dis},(i)} = - \frac{C_r}{C} \sum_{r'} I_{r',t}^{(i)}.
\end{equation}
However, since $I_{r,t}^{(i)}$ corresponds to the instantaneous
steady-state solution [see Eqs.~(\ref{eq:kineqin}) and (\ref{eq:obs})],
particle conservation implies $I_{L,t}^{(i)} + I_{R,t}^{(i)} = 0$ and
therefore $I_{r,t}^{\mathrm{dis},(i)}$ is exactly zero for any value of
$t$, as required. For the adiabatic correction to the displacement
current we have
\begin{equation}
 I_r^{\mathrm{dis},(a)} = C_r \dot{V}_r(t) - \frac{C_r}{C} \sum_{r'}
 C_{r'} \left[ \dot{V}_{r'}(t) + I_{r',t}^{(a)} \right].
\end{equation}
Here one notices that although this current can have finite values
during the driving cycle, its time-average over one period is zero. This
becomes evident since particle conservation yields $I_{L,t}^{(a)} +
I_{R,t}^{(a)} = -\frac{d}{dt}\braket{\hat{n}}_t^{(i)}$ such that the
above is a total time-derivative. 

In terms of the vector fields of the previous section, the displacement
current is related to an irrotational pseudo vector potential
\begin{equation}
 \mathbfcal{A}_{I_r^\mathrm{dis}} =  C_r \boldsymbol{\nabla} V_r - \frac{C_r}{C}
 \sum_{r'} C_{r'} \boldsymbol{\nabla} V_{r'} + \frac{C_r}{C} \boldsymbol{\nabla}
 \braket{\hat{n}}^{(i)},
\end{equation}
which can be imagined as a gauge function,
$\mathbfcal{A}_{I_r^\mathrm{dis}} = \boldsymbol{\nabla} \psi$, that
leaves the total pseudo magnetic field unchanged,
i.e. $\mathbfcal{B}_{I_r^\mathrm{tot}} = \boldsymbol{\nabla} \times
(\mathbfcal{A}_{I_r}+\boldsymbol{\nabla} \psi) = \boldsymbol{\nabla}
\times \mathbfcal{A}_{I_r}$. Since the pumped charge is due to this
pseudo magnetic field, we can focus in what follows only on tunneling
currents. 

Now that we have the formal expressions for the adiabatic charge and
spin currents, we can start with the analysis of the pumping for the
specific modulation of the gate and bias voltage [see
Eq.~(\ref{eq:voltages})] and fix the direction of the driving by taking
$\phi_g = -\pi/2$ and $\phi_b = \pi$ for which the adiabatic dc current
is maximal. Motivated by a previous study on interaction-induced
pumping,~\cite{Reckermann10} we describe in the following sections the
pumping of charge and spin making use of the vector fields 
of Eqs.~(\ref{eq:pseudoa}) and (\ref{eq:pseudob}). Since we are
interested in the effect of the interaction $U$ on the pumping, we first
consider the case $U=0$ as a reference. We then analyze the role of the
local interaction $U>0$ for $B=0$ and characterize the mechanism of
pumping. Finally, in Sec.~\ref{sec:Bfinite}, we extend the description
of the pumped charge and introduce the pumped spin by including an
external magnetic field $B>0$. 

\subsection{Non-interacting quantum dot}

For $U = 0, B \ge 0$ we observe that the charge current response
coefficients [see Eq. (\ref{eq:crescoeff})] are time-independent since
$\gamma_r = \beta_r =0$. Therefore, the adiabatic response is unaffected
by the change in the dot occupation during the full cycle. This means that
the same amount of charge that flows into the dot from the $r$-lead
during the {\em loading} part of the cycle, characterized by
$d\braket{\hat{n}}_t^{(i)}/dt > 0$, is returned to the same lead during 
the unloading. This constitutes a clear example in which the
loading/unloading symmetry is preserved after a complete cycle of the
driving. The analysis is also valid for the adiabatic spin current,
where the loading/unloading symmetry is manifested through the average
spin inside the dot. Therefore, although there is a finite charge and
spin current 
\begin{subequations}
\begin{align}
 I_{r,t}^{(a)} &= -\frac{\Gamma_r}{\Gamma}
 \frac{d}{dt}\braket{\hat{n}}_t^{(i)},\\
 J_{r,t}^{(a)} &= -\frac{\Gamma_r}{\Gamma}
 \frac{d}{dt}\braket{\hat{S}_z}_t^{(i)},
\end{align}
\end{subequations}
the total pumped charge and spin, obtained after integrating over the
full period, are exactly zero. In terms of the vector field introduced in
Eq.~(\ref{eq:pseudoa}), the constant response is then described as an
irrotational vector potential $\mathbfcal{A}_R = \boldsymbol{\nabla} \left[
\varphi_n^R \braket{\hat{n}}^{(i)} + \varphi_{S_z}^R
\braket{\hat{S}_z}^{(i)} \right]$, with $R$ either $I_r$ or $J_r$, whose
corresponding pseudo magnetic field $\mathbfcal{B}_{R} =
\boldsymbol{\nabla} \times \mathbfcal{A}_R$ is indeed zero. We emphasize
that this result depends on the particular choice of the driving
parameters. Other setups involving time-dependent barriers yield finite
pumping even in the $U=0$ limit since the response coefficients are
time-dependent and the loading/unloading symmetry is not necessarily
preserved.\cite{Splettstoesser06} 

\subsection{Interacting quantum dot at zero magnetic field}

We now focus on the effect of a finite Coulomb interaction $U \gg T >
\Gamma$ by considering $B=0$. Once the mechanism that generates the
pumping is understood, we extend, in Sec.~\ref{sec:Bfinite}, the
discussion to a finite magnetic field.

\begin{figure}[tbp]
\includegraphics[width=0.4\textwidth]{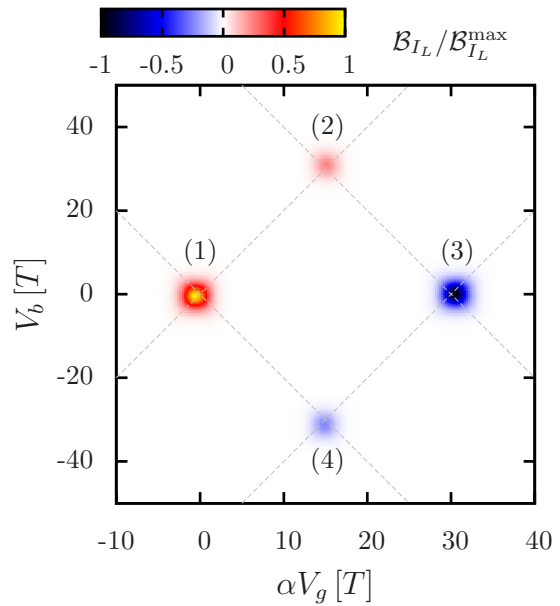}
\caption{(Color online) Stability diagram for the pumped charge:
 normalized pseudo magnetic field as function of the driving parameters
 $V_g$ and $V_b$, for $B = 0$. Grey dashed lines correspond to dot level
 resonance lines. The chosen Coulomb interaction and junction asymmetry
 are, respectively, $U=30T$ and $\lambda=0.25$.}
\label{fig:field}
\end{figure}

In Fig.~\ref{fig:field} we show the pseudo magnetic field of
Eq.~(\ref{eq:pseudob}) in terms of the driving parameters $V_g$ and
$V_b$ (from hereon called `stability diagram for the
pseudo magnetic field for the pumped charge'). This map is a convenient
tool in the characterization of the pumped charge (and later on the
pumped spin) for different regimes of the applied bias and arbitrary
asymmetry in the coupling to the leads. In this example, we show the
pseudo magnetic field $\mathcal{B}_{I_L}$ related to the adiabatic
charge current entering the left lead. As observed in
Ref.~\onlinecite{Reckermann10}, it displays four peaks of linear
dimension $\sim T$, located in the vicinity of the meeting point of two
resonance lines of the differential conductance, namely, at the corner
points of the regions of stable charge in the quantum dot. These lines
can be {\em approximated} by the dot-level resonance conditions $\mu_r =
\epsilon, \epsilon + U$, respectively, with $\epsilon = -\alpha V_g$ and
$r = L,R$ (grey dashed lines in the figure). The low-bias peaks [labeled
by (1) and (3) in Fig.~\ref{fig:field}] are dominant, while the
high-bias peaks [(2) and (4)], emerge only when we include an asymmetry
$\lambda = (\Gamma_L-\Gamma_R)/\Gamma$ between the two barriers. We
first investigate the regions in which $\mathcal{B}_{I_L}$ is zero and
then we separate the discussion of the peaks according to the applied
bias around which the driving takes place.

Since we consider $B = 0$, the singly occupied dot states are
degenerate, i.e. $\epsilon_{r\sigma}=\epsilon-\mu_r$ for $\sigma =
\uparrow,\downarrow$, such that $\braket{\hat{S}_z}_t^{(i)} = 0$ and the
adiabatic spin current completely vanishes. In this case, the adiabatic
charge current reduces to
\begin{subequations}
\begin{align}
 I_{r,t}^{(a)} &= -\frac{\Gamma_r+\gamma_r}{\Gamma+\gamma}
 \frac{d}{dt}\braket{\hat{n}}_t^{(i)},\\
 \braket{\hat{n}}_t^{(i)} &= 2 \frac{\sum_r
 \Gamma_r f(\epsilon_r/T)}{\Gamma+\gamma},
\end{align}
\label{eq:trccurnt}%
\end{subequations}
and admits the following interpretation: As soon as the driving
passes through a resonance line, the occupation in the quantum dot is
changed and generates a response current flowing from/into the
leads. The contribution flowing through the $r$-barrier is then given by
the ratio $(\Gamma_r+\gamma_r)/(\Gamma+\gamma)$ between the
charge relaxation rate relative to the $r$-lead,
i.e. $\Gamma_r+\gamma_r$, and the total relaxation rate $\Gamma+\gamma$
corresponding to the sum of the two leads. When taking the time-average
over a complete cycle, the occurrence of a dc component of the adiabatic
current is tied to an asymmetry between the loading and unloading
parts of the cycle. We can find two regimes in which this
condition is not fulfilled: (i) When the driving is far away from any
resonance line, the occupation in the dot remains constant and the
response is exactly zero. (ii) When the driving crosses a single
resonance line, the response is symmetric during the loading and
unloading. Therefore, the same amount of charge that enters the dot from
the $r$-lead during the loading returns to the same lead in the
unloading, such that the dc component of the adiabatic current is
zero. In terms of the pseudo magnetic field
\begin{equation}
\mathbfcal{B}_{I_r} = \boldsymbol{\nabla} \varphi_n^{I_r} \times
 \boldsymbol{\nabla}\braket{\hat{n}}^{(i)}, 
\label{eq:pseudobc}
\end{equation}
we observe that far away from any resonance line, the gradient of the
average charge is exponentially suppressed. On the other hand, when the
trajectory $\partial \Sigma$ traced by the driving parameters only
crosses a single resonance line, the response coefficient
$\varphi_n^{I_r}$ and the average charge $\braket{\hat{n}}^{(i)}$ depend
on the same effective parameter and the vectors in
Eq.~(\ref{eq:pseudobc}) become parallel to each other. In this sense,
adiabatic transport along a single resonance line can be understood as
single parameter pumping,~\cite{Kohler05} which is well known to give
zero contribution to the time-averaged current in the adiabatic
limit.\cite{Brouwer98}

In the remainder of this section, we analyze the discrete points around
which the pseudo magnetic field is nonzero and calculate the 2D
resonance shape of the related pumped charge. To this end, we first
notice that for the 
modulation we consider in Eq.~(\ref{eq:voltages}), it is convenient to
write the driving parameters as they enter in the arguments of the Fermi
function, i.e.
\begin{equation} 
\chi_r(t) = -\frac{\alpha V_g(t)}{T} \mp \frac{V_b(t)}{2T},
\label{eq:parameters}
\end{equation}
for $r=L,R$, respectively. Motivated by the location of the peaks in the
pseudo magnetic field, we investigate first the pumping at low bias,
characterized by the peaks marked by (1) and (3) in
Fig.~\ref{fig:field} and then we consider the high-bias peaks (2) and
(4). Once we know the behavior of these peaks, a complete description of
the stability diagram for the pumped charge can be obtained from the
symmetries of $\mathcal{B}_{I_r}$ with respect to a reflection of the
applied voltages (see Appendix \ref{app:symmetries}).

\subsubsection{Low-bias regime}
\label{sec:low-bias}

We now calculate the pseudo magnetic field for the low-bias peak (1)
around the point $(\alpha V_g,V_b) = (0,0)$. In this case, we take the
gradient of the response coefficient and the average charge in
Eq.~(\ref{eq:pseudobc}) with respect to the driving parameters of
Eq.~(\ref{eq:parameters}) and obtain
\begin{equation}
 \mathcal{B}_{I_L}^{(0,0)}(\boldsymbol\chi) \simeq \frac{\Gamma}
 {8(\Gamma+\gamma)^3} \frac{\Gamma_L\Gamma_R}{\cosh^2 \left(
 \frac{\chi_L}{2} \right) \cosh^2 \left( \frac{\chi_R}{2}\right)}, 
 \label{eq:bfield}
\end{equation}
with $\gamma \simeq \sum_r \Gamma_r f(\chi_r)$ and $\boldsymbol\chi =
(\chi_L,\chi_R)$. The superscript in $\mathcal{B}_{I_L}$ labels the
origin of coordinates with respect to which $\chi_L$ and $\chi_R$ is
measured. Notice here that the sign of the field is independent of
the coupling asymmetry $\lambda$. In particular, for the chosen
direction of the pumping cycle, the positive sign indicates that the
response current through the left lead is stronger during the unloading
part of the cycle. Additionally, the pseudo magnetic field decays
exponentially for $\chi \gg 1$, which, as we will show later, implies an
asymptotic value for the pumped charge when the enclosed area is larger
than the typical support of the field ($\sim 5T$).

\begin{figure}[tbp]
\includegraphics[width=0.4\textwidth]{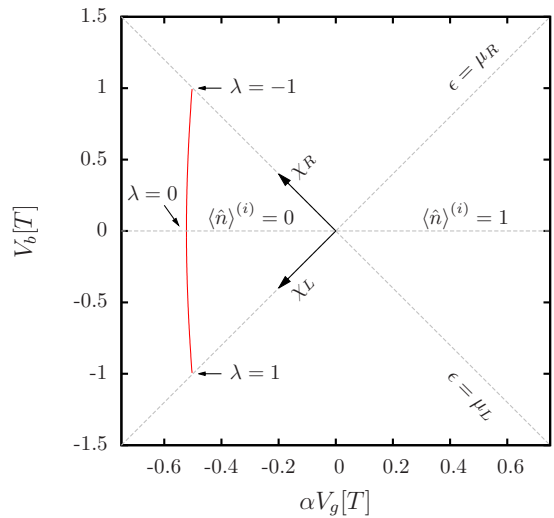}
\caption{Position of the low-bias peak (1) of the pseudo magnetic field
 as function of the junction asymmetry. The curve shown here, given by
 Eq.~(\ref{eq:maxb}), cannot be distinguished from the exact
 one.}
\label{fig:roots}
\end{figure}

Now that we know the specific shape of $\mathcal{B}_{I_L}$, we can
determine the condition at which the pumped charge is maximal, i.e. we
need to know the position $(\chi_L,\chi_R)$ where the pseudo magnetic
field reaches its maximum value. The exact position is obtained from the
roots of a quartic equation (see Appendix \ref{app:roots}), which can be
approximated through the interpolation between the maximum points for
$\lambda = 0$ and $\lambda = \pm 1$, i.e.
\begin{equation}
\chi_{L,R} \simeq \ln(1+\sqrt{3}) \lambda \frac{\lambda \pm 1}{2} 
 + \ln \left(\frac{1+\sqrt{33}}{4}\right)(1-\lambda^2).
\label{eq:maxb}
\end{equation}
In Fig.~\ref{fig:roots} we show the trajectory of the maximum value of
the field, whose location is given in Eq.~{\ref{eq:maxb}}, as the
junction asymmetry is swept over the whole range $-1<\lambda<1$. In the
symmetric case $\lambda = 0$, we observe that the peak is shifted with
respect to the charge degeneracy point $(\alpha V_g, V_b) = (0,0)$. Its
position is given by $\chi_L = \chi_R = \ln[(1+\sqrt{33})/4] \simeq
0.522$ or, in terms of the gate and bias voltage: $\alpha V_g
\simeq -0.522~T, V_b = 0$. The temperature dependent shift in the
pseudo magnetic field is similar to the known shift in the SET peak of
the linear conductance $G$.\cite{Bonet02} This last is
defined via the instantaneous current by
\begin{equation}
G = \left. \frac{dI_L^{(i)}}{dV_b} \right|_{V_b=0} =
\frac{\Gamma}{8T} \frac{1-\lambda^2}{1+f \left( \frac{\epsilon}{T}
\right)} \frac{1}{\cosh^2 \left(\frac{\epsilon}{2T}\right)}.
\end{equation} 
In addition to the broadening, the peak in $G$ shows a shift
$\Delta\epsilon = \tfrac{T}{2} \ln 2 \simeq 0.347~T$ which increases
linearly with $T$ as a consequence of the Coulomb interaction. This
effect is related to a change in the spin degeneracy of the ground state
when crossing the charge degeneracy point $(\alpha V_g, V_b) =
(0,0)$. The pseudo magnetic field can be expressed in terms of the
linear conductance through
\begin{subequations}
\begin{align}
\left. \mathcal{B}_{I_L}^{(0,0)} \right|_{V_b=0}  &=
 \frac{2}{1-\lambda^2}
 \frac{1}{1+f\left(\frac{\epsilon}{T}\right)}\left( \frac{T}{\Gamma}
 G \right)^2, \\
 &= \frac{1}{32}\frac{1-\lambda^2}{\left[1+f
 \left(\frac{\epsilon}{T}\right)\right]^3}\frac{1}{\cosh^4
 \left(\frac{\epsilon}{2T}\right)}, 
\end{align}
\end{subequations}
and though the origin of the peak shift in Eq.~(\ref{eq:maxb}) is the
same, the above level-dependent prefactor explains the different value
as compared to the one in $G$.

For a finite asymmetry in the tunnel couplings, the shift in the gate
voltage (see Fig.~\ref{fig:roots}) remains almost constant and the peak
moves vertically in the stability diagram. The extreme cases $\lambda =
-1$ and $\lambda = 1$ imply $(\alpha V_g,V_b) \simeq (-T/2,T)$ and
$(-T/2,-T)$ respectively, such that the peak sits close to the
resonance lines, as indicated by the arrows in Fig.~\ref{fig:roots}.

The extension of the above analysis to peak (3) in
Fig.~\ref{fig:field} is straightforward. In this case, we set the
origin of coordinates at the point $(\alpha V_g,V_b) = (U,0)$ and the
resulting pseudo magnetic field can then be related to the one of
Eq.~(\ref{eq:bfield}) by
\begin{equation}
\mathcal{B}_{I_L}^{(U,0)}(\chi_L,\chi_R) =
 -\mathcal{B}_{I_L}^{(0,0)}(-\chi_L, -\chi_R).
\label{eq:b12}
\end{equation}
The negative sign in the above equation indicates that now the loading
part dominates the cycle. Due to this sign, we now look at the points
that minimize the field here. According to the inversion of the sign in
the driving parameters, these points are the ones of Eq.~(\ref{eq:maxb})
but with opposite sign. Therefore, since we took the origin at $(\alpha
V_g,V_b) = (U,0)$, the peak (3) is shifted into the $N = 2$ Coulomb
diamond, such that for $\lambda = 0$, the peak is located at $\alpha V_g
\simeq U + 0.522~T$ and $V_b = 0$. Finally, when taking finite values
of $\lambda$, the shift of the position of the peak (3) as a function of
the bias is opposite to the one observed in the peak (1). In this sense,
for $\lambda \rightarrow -1$ the peak approaches to $(\alpha V_g,V_b) =
(U+T/2,-T)$ and when $\lambda \rightarrow 1$ it sits close to $(U+T/2,T)$.

\begin{figure}[tbp]
\includegraphics[width=0.4\textwidth]{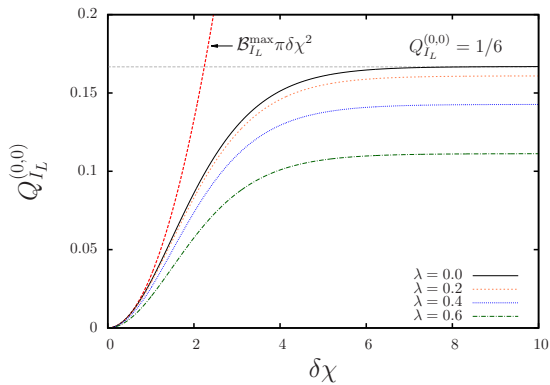}
\caption{(Color online) Pumped charge in the low bias regime
 as function of the modulation amplitude $\delta \chi$ for several
 values of the junction asymmetry. For $\lambda = 0$, the regimes of
 small and large driving amplitudes [see Eqs.~(\ref{eq:weakq}) and
 (\ref{eq:strongq})] are indicated by dashed red and gray lines
 respectively.} 
\label{fig:pcharge}
\end{figure}

We now calculate the {\em maximal pumped charge} as obtained
from Eqs.~(\ref{eq:QR})-(\ref{eq:pseudob}) when the working point is set
in the position where the peak is maximum [see Eq.~(\ref{eq:maxb})]. We
consider a circular trajectory $\partial \Sigma$ for the driving
parameters as defined in
Eq.~(\ref{eq:voltages}). The numerical evaluation of the pumped charge
$Q_{I_L}^{(0,0)}$ related to the peak (1) is depicted in
Fig.~\ref{fig:pcharge} as a function of the modulation amplitude
$\delta\chi$, i.e. the radius of the circle over which the field is
integrated. We observe that, regardless of the value of $\lambda$ or
$\delta \chi$, the sign of the pumped charge is fixed only by the
direction of the pumping cycle. As we will show in the next subsection,
this is not the case at high-bias, where the pumped charge shows a
strong dependence on $\lambda$.

For small driving amplitudes, i.e. $\delta\chi \ll 1$, the pumped charge
is proportional to the area $\Sigma$ encircled by the trajectory
$\partial \Sigma$ of the driving cycle, in agreement with
Refs.~\onlinecite{Brouwer98,Aleiner98}. For an arbitrary junction
asymmetry, this can be approximated by 
\begin{equation}
Q_{I_L}^{(0,0)} \simeq (1-\lambda^2)\mathcal{B}_{I_L}^\mathrm{max} \pi
 \delta\chi^2,
\label{eq:weakq}
\end{equation}
where $\mathcal{B}_{I_L}^\mathrm{max} \simeq 0.0106$ is the maximum
value of the peak for $\lambda =0$.

\begin{figure}[tbp]
\includegraphics[width=0.4\textwidth]{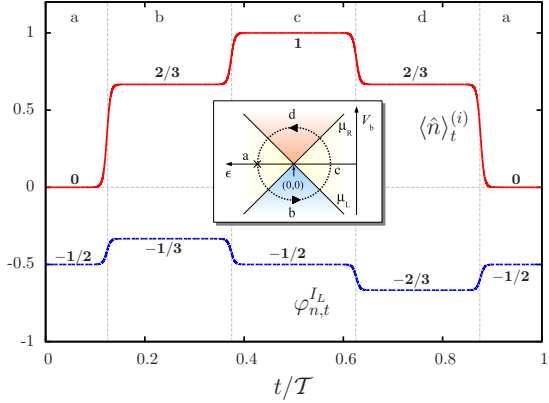}
\caption{(Color online) Time-resolved average charge (solid red) and
response coefficient (dashed blue) for $\lambda = 0$ and $\delta \chi
\gg 1$ during a cycle of the modulation. Inset: scheme of the explored
 regions of the stability diagram during the pumping cycle around the
 working point $(\alpha \bar{V}_g,\bar{V}_b)=(0,0)$.}
\label{fig:onesixth}
\end{figure}

When increasing the amplitude, the driving parameters start exploring
regions which are away from the crossing point of the two resonance
lines and the pumped charge no longer follows the above relation. In
this large amplitude (but still adiabatic) driving regime, characterized
by $\delta\chi \gtrsim 1$, it is important to remark that the adiabaticity
condition $\Omega \delta \chi \ll T$ is still preserved, since we can
always take arbitrary small values for the modulation frequency without
affecting $Q_{I_L}$. When evaluating the pseudo magnetic field in
Eq.~(\ref{eq:pseudobc}) for $\chi_r \gg 1$, the exponential decay
implies an asymptotic value for the pumped charge, as shown in
Fig.~\ref{fig:pcharge}. In this case, the particular choice of the
working point and the specific shape of the trajectory in the parameter
space become irrelevant as far as the peak is fully contained in
$\Sigma$. The pumped charge then saturates at\cite{note2}
\begin{equation}
 Q_{I_L}^{(0,0)} \simeq \frac{3}{2} \frac{1-\lambda^2}{9-\lambda^2}.
\label{eq:strongq}
\end{equation}
It shows a quadratic dependence on $\lambda$ for $\lambda \ll 1$, in
agreement with Fig.~\ref{fig:pcharge}. The maximum value of the pumped
charge, corresponding to $\lambda =0$, is
\begin{equation}
Q_{I_L}^{(0,0)}|_\mathrm{max} = \frac{1}{6},
\end{equation}
in units of the electronic charge. To explain this particular value and
to illustrate the mechanism of pumping at low bias we show, in
Fig.~\ref{fig:onesixth}, the time-resolved average charge and response
coefficient during a cycle of the modulation around the working point 
$(\alpha \bar{V}_g,\bar{V}_b) = (0,0)$. We divide the cycle into four
steps corresponding to the different regions of the stability diagram
visited by the driving parameters. Here we consider a symmetric coupling
to the leads ($\lambda = 0$). The four regions are:

\begin{enumerate}[a)]
\item Turning point (meaning that the dot level takes its maximum or
      minimum value) above the chemical potentials: $\epsilon >
      \mu_r$. The $\gamma_r$ factors are exponentially suppressed and
      $\varphi_{n}^{I_r} = -1/2$.

\item Going down between the chemical potentials: $\mu_L < \epsilon
      < \mu_R$. Here $\gamma_L \simeq 0$ and $\gamma_R \simeq \Gamma/2$,
      such that $\varphi_{n}^{I_L} =  -1/3$ and $\varphi_{n}^{I_R} =
      -2/3$.

\item Turning point below: $\epsilon < \mu_r$. Here $\gamma_r \simeq
      \Gamma/2$ and the same response coefficients are obtained. As in
      a), these are $-1/2$.

\item Going up through $\mu_R < \epsilon < \mu_L$, the asymmetric
      situation observed in b) is reversed and we obtain
      $\varphi_{n}^{I_L} = -2/3$ and $\varphi_{n}^{I_R} = -1/3$.
\end{enumerate}

To estimate the pumped charge, we consider the time-integral of
the current given in Eq.~(\ref{eq:trccurnt}) noting that, in the large
amplitude driving regime, the time-derivative of
$\braket{\hat{n}}_t^{(i)}$ is only nonzero when crossing a resonance
line. This last derives from the blocking of certain transitions by the
Coulomb interaction. Therefore, we can approximate $d
\braket{\hat{n}}_t^{(i)}/dt$ through the difference between the
asymptotic values obtained at each side of the resonance, while for the
response coefficient we take the average between its asymptotic values
before and after the crossing, i.e. $Q_{I_L}^{(0,0)} \simeq
-5/18-5/36+7/18+7/36 = 1/6$. This corresponds to the maximum possible
value of the pumped charge when modulating the gate and bias voltage. We
observe that for this particular modulation the total pumped charge is 
not quantized. Such quantization of $Q_{I_L}^{(0,0)}$ would demand the
modulation of an additional parameter (e.g. the tunnel
barriers)\cite{Battista11} and is not what we address here. In general,
the (measurable) plateau value that is reached does, however, provide
information about the tunnel coupling asymmetry. This is similar to the
use of noise values in the SET regime.\cite{Thielmann03} 

To complete the analysis of pumping at low bias, we mention that the
pumped charge $Q_{I_L}^{(U,0)}$ related to the peak (3) around the point
$(\alpha V_g, V_b) = (U,0)$ reads 
\begin{equation}
Q_{I_L}^{(U,0)} = - Q_{I_L}^{(0,0)},
\end{equation}
where we used the antisymmetric shape of $\mathcal{B}_{I_L}$ along the
particle-hole symmetry point $(\alpha V_g,V_b) = (U/2,0)$ [see
Eq.~(\ref{eq:b12}) and Appendix \ref{app:symmetries}].

\subsubsection{High-bias regime}
\label{sec:high-bias}

Now we extend the above discussion to the region around a large static
bias $V_b \sim U$, i.e. the peak (2) in Fig.~\ref{fig:field}. We
evaluate the pseudo magnetic field and the resulting pumped charge when
encircling the crossing point [cf. Eq.~(\ref{eq:voltages})] $(\alpha
V_g, V_b) = (U/2,U)$, which is taken as the new origin of
coordinates for the voltages. In this regime, the resulting pseudo
magnetic field can be written in terms of the low-bias field as
\begin{equation}
 \mathcal{B}_{I_L}^{(U/2,U)}(\chi_L,\chi_R) = \lambda \,
  \mathcal{B}_{I_L}^{(0,0)}(-\chi_L,\chi_R).
\label{eq:hbpseudob}
\end{equation}
This relation between low- and high-bias field is a central result of
this paper. As a direct consequence of the $\lambda$ prefactor, we
observe that since $|\lambda|<1$, the magnitude of the peak (2) is
always smaller than the one of peak (1) and its sign is uniquely
determined by the sign of $\lambda$. For any two modulation curves of
the same shape and direction, centered around these points and symmetric
with respect to the $\chi_R$ axis, a change of variables allows us to write
\begin{equation}
\iint_\Sigma dS \mathcal{B}_{I_L}^{(U/2,U)}(\boldsymbol{\chi})
 = \lambda \iint_{\Sigma} dS \mathcal{B}_{I_L}^{(0,0)}(\boldsymbol{\chi}),
\end{equation}
such that we can calculate $Q_{I_L}^{(U/2,U)}$ in terms of the pumped
charge at low bias
\begin{equation}
Q_{I_L}^{(U/2,U)} = \lambda \, Q_{I_L}^{(0,0)}.
\end{equation}
As noted in Ref.~\onlinecite{Reckermann10}, the mere presence of a
pumped charge in the high-bias regime indicates an asymmetric coupling
to the leads. Since for the chosen modulation the sign of
$Q_{I_L}^{(0,0)}$ is always positive, the sign of $Q_{I_L}^{(U/2,U)}$
could be used as a quick test to determine which one of the two leads is
dominating the transport. For a direct quantitative estimation of
$\lambda$, one simply divides the pumped charge at the different bias
regimes:
\begin{equation}
\frac{\Gamma_L-\Gamma_R}{\Gamma_L+\Gamma_R} =
 \frac{Q_{I_L}^{(U/2,U)}}{Q_{I_L}^{(0,0)}}.
\end{equation}
In particular, this is convenient in the regime of large driving
amplitudes, where the pumped charge is not affected by the precise
details of the trajectory. 

As compared to the low-bias peak (1), the $\lambda$ prefactor tells us
that the loading/unloading symmetry can no longer be broken for
symmetric barriers, i.e. $\lambda = 0$. In fact, for this particular
case, although there is a change in the response coefficient along the
pumping cycle, it has the same time-dependence as the average charge,
i.e.
\begin{equation}
\varphi_{n,t}^{I_L}=\frac{\braket{\hat{n}}_t^{(i)}}{2}-1,
\end{equation}
and the adiabatic charge current is a total time-derivative. Therefore,
a finite pseudo magnetic field is now a cooperative effect of a change
in the response coefficient and the junction asymmetry: A finite
$\lambda$ is required in order to have non-parallel gradients in
Eq.~(\ref{eq:pseudobc}). 

Eq.~\ref{eq:hbpseudob} is also useful in that it allows us to determine
the maximal pumped charge working point based entirely on the low-bias
feature. According to Eq.~(\ref{eq:hbpseudob}), the dependence on
$\lambda$ of this point follows the same condition as in the low-bias
regime, except for the sign inversion of $\chi_L$ and the shifted origin
of coordinates. Therefore, starting from the position of the peak (1) in
Fig.~\ref{fig:field}, we can determine the position of the peak (2) by
performing first a reflection at the resonance line $\epsilon =
\mu_L$ and then a translation by $(U/2,U)$. Although
for symmetric junctions ($\lambda = 0$) there is no peak, its position
would be shifted in the bias voltage by $\Delta V_b = 2
\ln[(1+\sqrt{33})/4]~T$ with respect to the crossing of the dot-level
resonance lines at $(U/2,U)$. As soon as we increase $\lambda$ from $0$,
the peak emerges and moves almost horizontally (i.e. along the gate
voltage axis) towards the $\epsilon = \mu_R$ resonance line whereas for
negative values of $\lambda$ the peak moves towards the $\epsilon =
\mu_L$ line. 

Finally, the above analysis can be transferred to the peak (4) by using
the bias voltage symmetry discussed in Appendix \ref{app:symmetries}. In
this case, the pseudo magnetic field can be written in terms of the
low-bias field as
\begin{equation}
\mathcal{B}_{I_L}^{(U/2,-U)}(\chi_L,\chi_R) =
-\lambda \mathcal{B}_{I_L}^{(0,0)}(\chi_L,-\chi_R),
\end{equation}
such that the pumped charge at this working point 
\begin{equation}
Q_{I_L}^{(U/2,-U)} = -\lambda Q_{I_L}^{(0,0)},
\end{equation}
provides an additional and independent quantitative measurement of the
junction asymmetry. This can be used as a cross check on experimental
results.

\subsection{Finite external magnetic field}
\label{sec:Bfinite}

We now include a finite external magnetic field $B \gg T$.~\cite{note3}
In this situation, in addition to the adiabatic charge current
$I_{r,t}^{(a)}$, a nonzero adiabatic spin current $J_{r,t}^{(a)}$ also
flows through the dot.

\begin{figure*}[ht!]
 \begin{center}
  \subfigure{\includegraphics[width=0.25\textwidth]{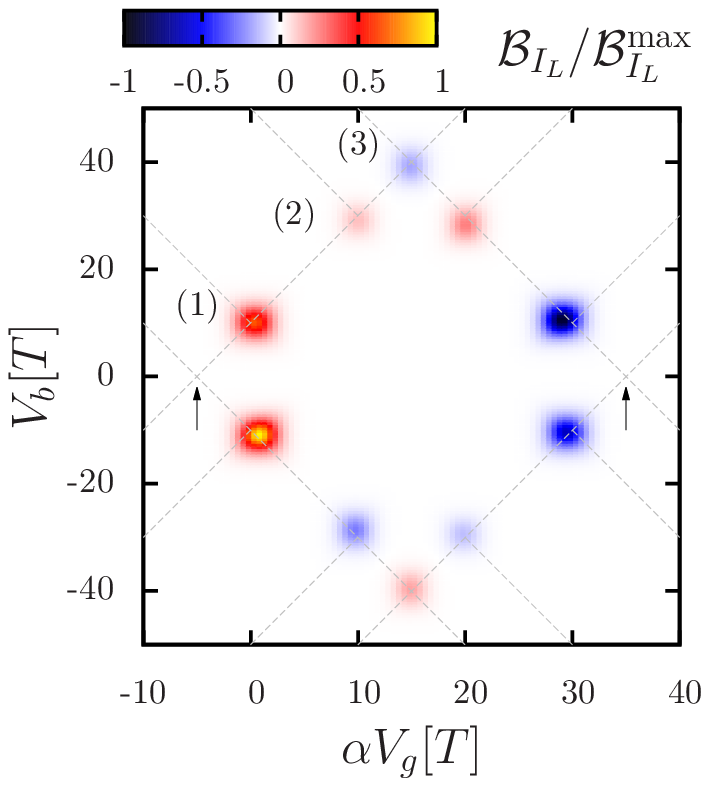}}
  \subfigure{\includegraphics[width=0.25\textwidth]{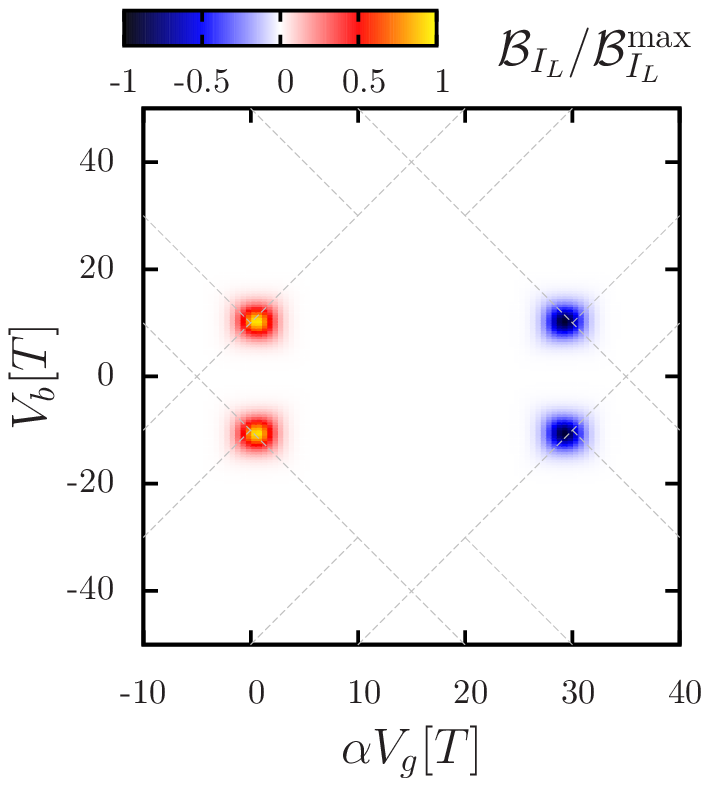}}
  \subfigure{\includegraphics[width=0.25\textwidth]{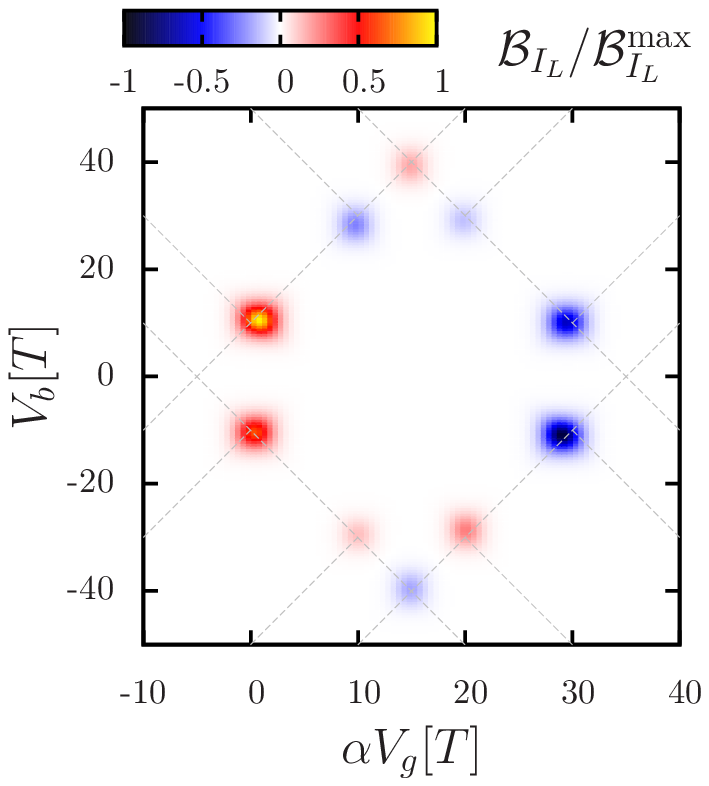}}\\
  \subfigure{\includegraphics[width=0.25\textwidth]{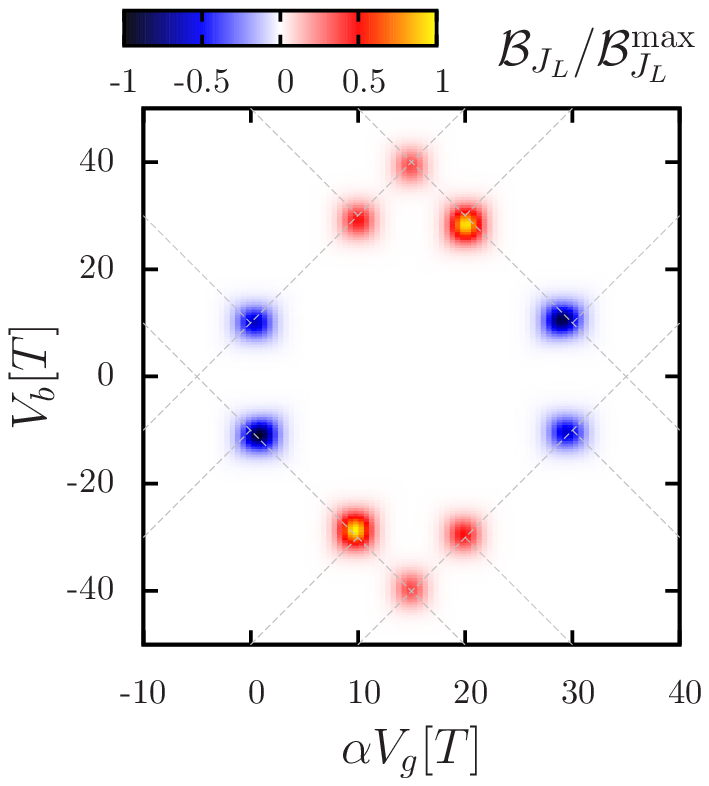}}
  \subfigure{\includegraphics[width=0.25\textwidth]{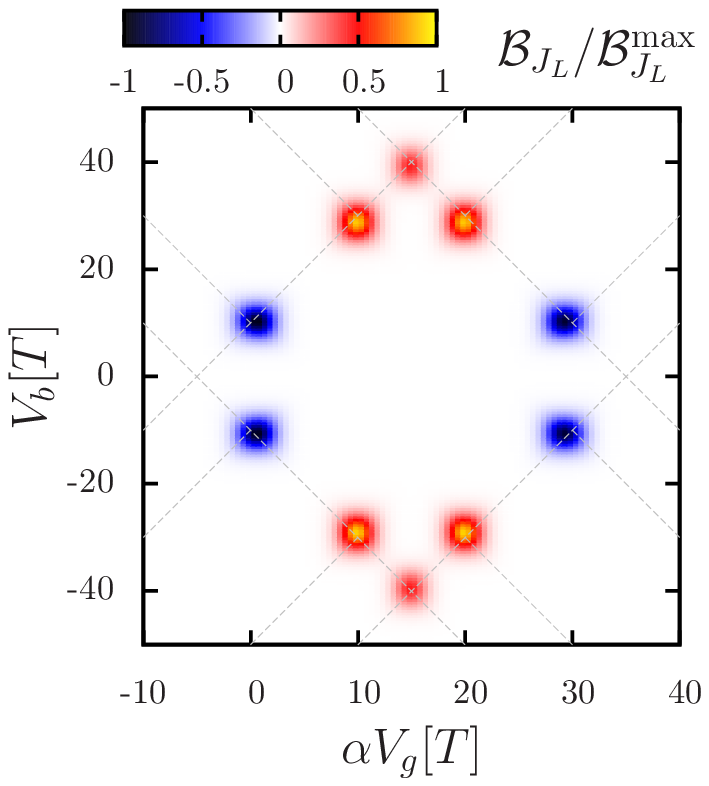}}
  \subfigure{\includegraphics[width=0.25\textwidth]{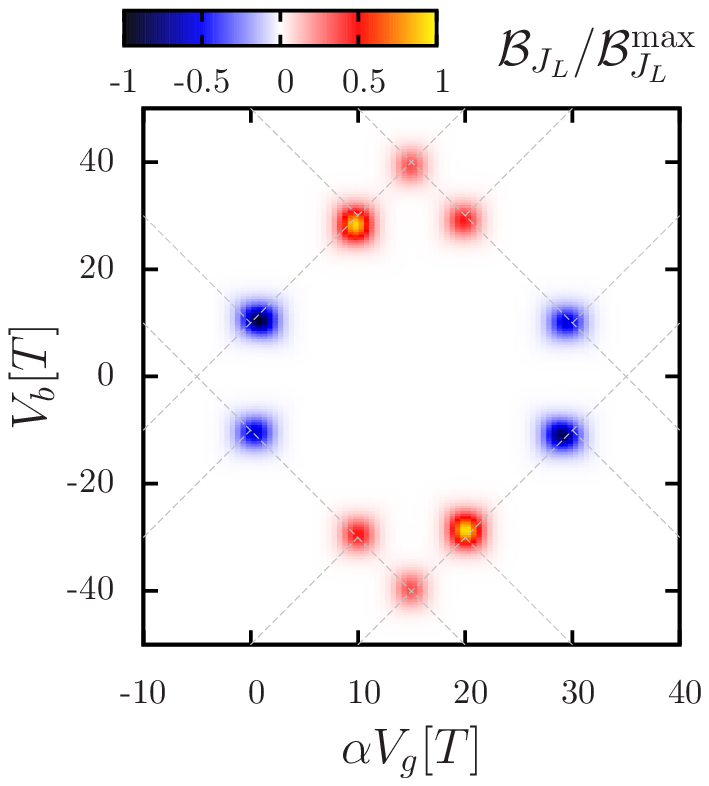}}
 \end{center}
 \caption{(Color online) Normalized pseudo magnetic fields for the
 pumped charge (top panels) and pumped spin (bottom panels) for junction
 asymmetries $\lambda = -0.25$ (left), $\lambda = 0$ (center) and
 $\lambda = 0.25$ (right). The chosen interaction energy is $U = 30T$
 and the external magnetic field is $B = 10T$.}
 \label{fig:pseudob1}
\end{figure*}

In Fig.~\ref{fig:pseudob1} we show stability diagrams for the pseudo
magnetic fields for the pumped charge (top panels) and the
pumped spin (bottom panels) for different junction asymmetries. The
external magnetic field $B = 10~T$, where $T$ is the 
thermal energy, now splits the resonances of Fig.~\ref{fig:field} into
further well-separated peaks. The pseudo magnetic field is nonzero only
around the meeting points of two crossing resonance lines. As in the
case $B = 0$, this is again because the loading/unloading symmetry is
preserved when the driving is far away from any resonance line or when it
only crosses a single resonance line. A simple inspection of
Fig.~\ref{fig:pseudob1} 
shows that, regardless of the value of $\lambda$, there is no peak in
the crossing point at zero bias (black arrows in the upper left
panel). In Ref.~\onlinecite{Reckermann10}, the absence of a peak was 
related to the lifting of the spin degeneracy in the states with single
occupation. Additionally, a strong dependence on $\lambda$ in the
high-bias peaks [(2) and (3) in Fig.~\ref{fig:pseudob1}] of
$\mathcal{B}_{I_L}$ is observed.\cite{Reckermann10} In contrast, the 
peaks of $\mathcal{B}_{J_L}$ are almost unaffected by $\lambda$. The
difference between charge and spin current is particularly strong for
$\lambda = 0$, where we find {\em pure} spin pumping in the high-bias
regime: the pumped charge peaks (top center panel) vanish exactly while
the ones related to the pumped spin (bottom center panel) remain
finite. 

To understand the above features of the pseudo magnetic fields and how
they affect the pumping, we consider first the absence of a peak around
$(\alpha V_g, V_b) = (-B/2,0)$. In this regime of the driving
parameters, the charge in the quantum dot is spin polarized, such that
the average charge and spin simplify to $\braket{\hat{n}}_t^{(i)} =
p_{\uparrow,t}^{(i)}$ and $\braket{\hat{S}_z}_t^{(i)} =
p_{\uparrow,t}^{(i)}/2$, respectively. Therefore, the adiabatic currents
of Eqs.~(\ref{eq:ccurnt_ad}) and (\ref{eq:scurnt_ad}) simplify to
\begin{subequations}
\begin{align}
I_{L,t}^{(a)} &= -\frac{\Gamma_L}{\Gamma} \frac{d}{dt}
 \braket{\hat{n}}_t^{(i)},\\ 
J_{L,t}^{(a)} &= \frac{1}{2} I_{L,t}^{(a)}.
\end{align}
\end{subequations}
Since in this regime there is a single available transition, namely,
$\ket{0} \leftrightarrow \ket{\uparrow}$, the relative rate at which the
dot is loaded has to be the same as the one
during the unloading. In the language of the vector fields, the vector
potential $\mathbfcal{A}_R$ associated with these currents is
irrotational, such that integration over the closed trajectory yields
zero pumped charge. As compared to the $B = 0$ result at low bias (see
Sec.~\ref{sec:low-bias}), a finite pumped charge requires not only a
modulation encircling the meeting point of two resonance lines but also
a change in the spin degeneracy of the ground state.

\begin{figure}[tbp]
\includegraphics[width=0.4\textwidth]{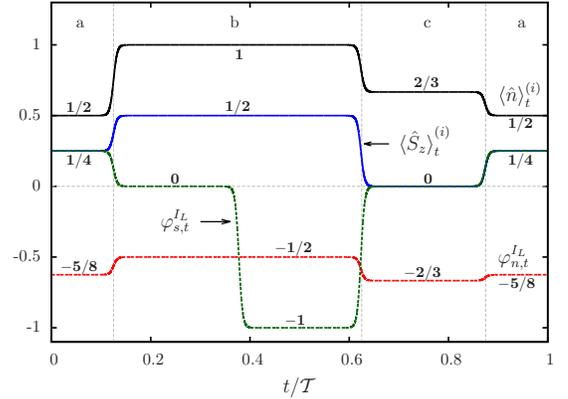}
\caption{(Color online) Time-resolved average charge (solid black) and
 spin (solid blue) together with the charge current response
 coefficients $\varphi_{n,t}^{I_L}$ (dashed red) and
 $\varphi_{s,t}^{I_L}$ (dashed green) for $\lambda = 0$ and $\delta \chi
 \gg 1$ during a cycle of the modulation around the working point
 $(0,B)$.}
\label{fig:onethird}
\end{figure}

We now investigate the regions in which the pseudo magnetic field is
nonzero. We consider first the peak labeled by (1) in
Fig.~\ref{fig:pseudob1}. Here, the two transitions $\ket{0}
\leftrightarrow \ket{\uparrow}$ and $\ket{0} \leftrightarrow
\ket{\downarrow}$ are enabled by the bias window $\mu_L-\mu_R$ such
that the spin degeneracy in the $N=1$ charge block is effectively
recovered and the loading/unloading symmetry is again broken. To
calculate the pseudo magnetic fields associated with the charge
($\mathcal{B}_{I_L}$) and spin ($\mathcal{B}_{J_L}$) currents, we set
the origin of coordinates at the crossing point $(\alpha V_g,V_b) =
(0,B)$. Although the explicit expression for the fields is cumbersome,
these show the simple relation
\begin{equation}
\mathcal{B}_{J_L}^{(0,B)}(\boldsymbol{\chi}) = -
 \frac{1}{2}\mathcal{B}_{I_L}^{(0,B)}(\boldsymbol{\chi}),  
\end{equation}
for arbitrary junction asymmetries. By calculating the corresponding
pumped charge and spin, and noticing that these follow the same relation
as the fields, we can describe the spin resolved pumped charge through
the definitions
\begin{subequations}
\begin{align}
Q_{I_L^\uparrow} = \frac{Q_{I_L}+2Q_{J_L}}{2},\\
Q_{I_L^\downarrow} = \frac{Q_{I_L}-2Q_{J_L}}{2}.
\end{align}
\label{eq:spinq}%
\end{subequations}
Since in this regime $Q_{J_L}^{(0,B)}=-Q_{I_L}^{(0,B)}/2$, the pumped
charge is purely given by spin $\downarrow$ carriers. To estimate the
dependence on $\lambda$, we calculate $Q_{I_L}^{(0,B)}$ for a large
driving amplitude and obtain 
\begin{equation}
Q_{I_L}^{(0,B)} = \frac{1+\lambda}{3+\lambda}.
\label{eq:onethird}
\end{equation}
Notice that, for the chosen modulation, the sign of the pumped charge is
always positive, regardless of the particular value of $\lambda$. Since
no charge is accumulated after one period of the modulation,
i.e. $Q_{I_L}^{(0,B)} + 
Q_{I_R}^{(0,B)} = 0$, the charge is pumped from the right lead to the
left lead. 
In this case, the breaking of the loading/unloading symmetry in both the
charge and spin in the dot is mainly due to the change in the number of
available transitions. This affects the value of the response
coefficients in such a way that the total amount of charge leaving the
left lead during the loading part of the cycle is smaller than the one
entering the same lead during the unloading. This is illustrated in
Fig.~\ref{fig:onethird}, for the case $\lambda = 0$ and $\delta\chi \gg
1$, where we show the average charge and spin in the dot together with
the corresponding response coefficients. According to
Eq.~(\ref{eq:onethird}), the pumped charge for $\lambda = 0$ corresponds
to $1/3$ of the electronic charge. This limit can be understood from the
asymptotic values shown in Fig.~\ref{fig:onethird}, in a similar way as
we did in the previous section for $Q_{I_L}^{(0,0)}$. The difference 
now is that the pumped charge is not only subjected to a change in the
average charge of the dot, but also to the variation of the average
spin. In particular, the loading (unloading) of the spin does not
necessarily correlate with the loading (unloading) of the
charge. Therefore we must distinguish the two contributions. The 
estimation of the time-derivatives of the average charge and spin,
together with the average value of the response coefficients each time
the driving parameters cross a resonance line (gray dashed lines in
Fig.~\ref{fig:onethird}) yields
\begin{equation}
Q_{I_L}^{(0,B)} = 1/48 + 15/48 = 1/3,
\label{eq:onethird1}
\end{equation}
with the first term related to the variation of the average charge and
the second to the average spin.

The above discussed value $1/3$ for the maximal pumped charge can be
increased or decreased depending on the sign of the junction
asymmetry [see Eq.~(\ref{eq:onethird})]. For positive $\lambda$, the
maximal pumped charge increases from $1/3$ to $1/2$. However, in the
extreme case $\lambda \simeq 1$ the (negative) peak (2) approaches the
region of integration and its contribution can no longer be
disregarded. 

We now consider the pumped charge and spin in the high-bias limit,
characterized by the peaks (2) and (3) in Fig.~\ref{fig:pseudob1}. To
keep the notation simple, we now use the index (2) to indicate that the
origin of coordinates is set at the point $(\alpha V_g,V_b) =
((U-B)/2,U)$. In the regime of large driving amplitude we obtain 
\begin{subequations}
\begin{align}
 Q_{J_L}^{(2)} &= \frac{1}{2} \frac{(1+\lambda)^2(3-\lambda)}
 {(3+\lambda)(3+\lambda^2)},\\ 
 Q_{I_L}^{(2)} &= -2 \lambda Q_{J_L}^{(2)}.
\end{align}
\end{subequations}
Therefore, since the pumped spin is always positive, the resulting
pumped charge changes its sign in the symmetry point $\lambda =
0$. Notice that the regime of validity of the above limit includes
$|\lambda| < 1$. For $\lambda \simeq 1$ we should consider the
contribution from the peak (1), such that both the pumped charge and
spin go to zero. By using Eq.~(\ref{eq:spinq}), the spin resolved
current shows to be decomposed into contributions from spin $\uparrow$
carriers flowing from the right lead to the left lead and spin
$\downarrow$ carriers flowing in the opposite direction. The ratio
between these two contributions depends on $\lambda$ via
\begin{equation}
\left|
 \frac{Q_{I_L^\uparrow}^{(2)}}{Q_{I_L^\downarrow}^{(2)}}\right|
= \frac{1-\lambda}{1+\lambda},
\end{equation}
such that for $\lambda < 0$ ($>0$) the transport is dominated by spin
$\uparrow$ ($\downarrow$) carriers. Remarkably, for a symmetrically
coupled dot we obtain pure spin pumping, since the two contributions are
exactly opposite and therefore the pumped charge is zero. In
this case, the limit $\delta\chi \gg 1$ yields
\begin{equation}
Q_{J_L}^{(2)} = \frac{1}{6},
\end{equation}
which can again be understood in terms of the asymptotic values of the
response coefficients and the instantaneous average charge and spin as
we previously demonstrated for $Q_{I_L}^{(0,B)}$ [see
Eq.~(\ref{eq:onethird1})]. 

Finally, we analyze pumping for large driving amplitudes around the peak
(3), given by the crossing of the resonance lines at $(\alpha V_g,
V_b) = (U/2,U+B)$. Now the relation between the pumped charge and
spin is non-trivial:
\begin{subequations}
\begin{align}
 Q_{I_L}^{(3)} &= \lambda \frac{1-\lambda^2}{3+\lambda^2},\\
 Q_{J_L}^{(3)} &= \frac{1}{4} (1-\lambda^2)
  \frac{3+4\lambda^2+\lambda^4}{(3+\lambda^2)^2}.
\end{align}
\end{subequations}
It implies a positive sign for the pumped spin while the pumped charge
again changes its sign with the sign of $\lambda$. Despite this, the
spin resolved pumped charge, calculated by Eq.~(\ref{eq:spinq}), obeys
$Q_{I_L^\downarrow}^{(3)}(\lambda) = -
Q_{I_L^\uparrow}^{(3)}(-\lambda)$. As in the previously studied regime
around the peak (2), symmetric coupling to the leads yields a pure
pumped spin, which in this case is
\begin{equation}
Q_{J_L}^{(3)} = \frac{1}{12}.
\end{equation}
The difference with the situation around the peak (2) is that now, for
$\lambda < 0$ ($>0$), the transport is dominated by $\downarrow$
($\uparrow$) carriers.

Finally, we note that the above discussion of peaks (1)-(3) can be
extended to all remaining peaks in Fig.~\ref{fig:pseudob1} by using
general gate- and bias-voltage symmetries of the problem in an external
magnetic field which are presented in Appendix~\ref{app:symmetries}.

\section{Conclusions}
\label{sec:summary}

We investigated adiabatic charge and spin pumping through
an interacting quantum dot driven out of equilibrium by a non-linear
bias voltage and time-dependent parameter modulations. We showed
that, regardless of the specific modulation, the time-resolved adiabatic
charge and spin currents can be interpreted as the response to a
perturbation in the instantaneous average charge and spin due to the
variation of the driving parameters. This allowed us to identify the
charge and spin emissivities to the leads in the context of interacting
systems in the non-linear transport regime.

For the specific case of a modulation of the gate and bias voltage, we
discussed the conditions for interaction-induced pumping in terms of the
properties of vector fields associated to the adiabatic charge and spin
currents. We observed that the Coulomb interaction is crucial since it
gives the rotational contribution to the pseudo vector potential that
cannot be gauged away: It generates a nonzero pseudo magnetic field and
in consequence a finite pumped charge. For a single-level quantum dot,
we explored the stability diagram associated to this vector field for
arbitrary bias and junction asymmetry. The shape of the pseudo magnetic
field reflects the two-parameter condition required for adiabatic
pumping, such that it shows a maximum whenever {\em two} lines of the
usual $dI/dV$ stability diagram meet. The analytic expressions for the
pseudo magnetic field and the pumped charge enable detailed fitting of
experimental results. For low bias voltages, the pumping mechanism is
dominated by the change in the response coefficient when exploring
different regions of the pumping cycle. In constrast, in the high-bias
regime, the finite pumped charge is generated by the cooperative effect
of the above mechanism and the junction asymmetry. This allows for a
direct quantitative determination of the junction asymmetry by two
measurements of the pumped charge.

The role of the external magnetic field was found to be twofold: First,
it restores at low bias the loading/unloading symmetry previously broken
by the local interaction in combination with the spin-degeneracy. This is
evidenced by a suppression of the pumped charge at zero bias. Second, in
addition to the pumped charge, a pumped spin arises once the spin
degeneracy is effectively recovered through an applied bias. In particular,
the weak dependence of the pumped spin on the junction asymmetry allows
for pure spin pumping in the high-bias regime.

\section*{Acknowledgments}

We acknowledge helpful discussions with M. B\"{u}ttiker, F. Haupt,
R.-P. Riwar and L.~E.~F. Foa Torres, and financial support from the DFG
under Contract No. SPP-1243 and the Ministry of Innovation NRW.

\appendix

\section{Applied voltage symmetries}
\label{app:symmetries}

In order to completely characterize the adiabatic transport of charge
and spin along the full stability diagram, we derive the reflection
symmetries of the pseudo magnetic fields $\mathcal{B}_{I_L}$ and
$\mathcal{B}_{J_L}$. Specifically, we study the behavior of
$\mathcal{B}_R$ with respect to: (i) the reversal of the bias
voltage $V_b \rightarrow -V_b$ and, (ii) the reversal of the
gate voltage $\alpha V_g \rightarrow U/2 - \alpha V_g$. To simplify the
notation, we use $v = V_b/2$. These symmetries derive from the
interchange of the two electrodes and the particle-hole symmetry of a
single interacting level. Although not exact anymore for multi-level
quantum-dots, we expect similar qualitative correspondences to hold in
general. 

\subsection{Bias voltage reversal}
We write the coupling strength in terms of the junction asymmetry by
$\Gamma_r(\lambda) = \Gamma (1+\alpha_r \lambda)/2$, with $\alpha_r =
\pm$ for $r=L,R$ respectively. The factors in Eq.~({\ref{eq:rates}})
then obey the following relations
\begin{subequations}
 \begin{align}
  \gamma_r(\epsilon,-v,\lambda) &= \gamma_{\bar{r}}(\epsilon,v,-\lambda),\\
  \beta_r(\epsilon,-v,\lambda)  &= \beta_{\bar{r}}(\epsilon,v,-\lambda),  
 \end{align}
\end{subequations}
where $\bar{r} = R (L)$ for $r = L (R)$ and $\epsilon = -\alpha V_g$. By
plugging these into the definitions of the charge current response
coefficients [see Eq.~(\ref{eq:crescoeff})] we obtain 
\begin{subequations}
 \begin{align}
  \varphi_n^{I_L}(\epsilon,-v,\lambda) &=
  -1-\varphi_n^{I_L}(\epsilon,v,-\lambda),\\ 
  \varphi_{S_z}^{I_L}(\epsilon,-v,\lambda) &=
  -\varphi_{S_z}^{I_L}(\epsilon,v,-\lambda),  
 \end{align}
\end{subequations}
and for the spin current response coefficients of
Eq.~(\ref{eq:srescoeff}) we have
\begin{subequations}
\begin{align}
\varphi_n^{J_L}(\epsilon,-v,\lambda) &= -\varphi_n^{J_L}(\epsilon,v,-\lambda),\\ 
\varphi_{S_z}^{J_L}(\epsilon,-v,\lambda) &= -1-\varphi_{S_z}^{J_L}(\epsilon,v,-\lambda).
\end{align}
\end{subequations}
Now we repeat this analysis for the average charge and spin. The
explicit expressions of these two quantities for arbitrary $U$ and $B$
can be calculated from the instantaneous occupation probabilities
$\mathbf{p}_t^{(i)}$ obtained as the solution to
Eq.~(\ref{eq:kineqin}). These write as follows
\begin{subequations}
\begin{align}
\braket{\hat{n}}^{(i)} &= \frac{\sum_{r\sigma}\Gamma_r
 f(\epsilon_{r\bar{\sigma}}) (\Gamma-2\gamma_\sigma)}
 {\Gamma^2-\gamma^2+\beta^2},\\
\braket{\hat{S}_z}^{(i)} &= \frac{1}{2} \frac{\sum_{rr'\sigma} \Gamma_r
 \Gamma_{r'} f(\epsilon_{r\sigma})
 f^{-}(\epsilon_{r\bar{\sigma}}+U)}{\Gamma^2-\gamma^2+\beta^2},
\end{align}
\end{subequations}
where $\bar{\sigma} = -\sigma$, $f^{-}(\omega) = f(-\omega)$, with
$f(\omega) = [1+\exp(\omega/T)]^{-1}$ and $\gamma_\sigma = \sum_r
\Gamma_r/2 [f(\epsilon_{r\sigma}) - f(\epsilon_{r\sigma}+U)]$. For these
averages, a change in the sign of the bias voltage is equivalent to an
inversion of the tunnel barriers, i.e.
\begin{subequations}
\begin{align}
\braket{\hat{n}}^{(i)}(\epsilon,-v,\lambda) &=
 \braket{\hat{n}}^{(i)}(\epsilon,v,-\lambda)\\  
\braket{\hat{S}_z}^{(i)}(\epsilon,-v,\lambda) &=
 \braket{\hat{S}_z}^{(i)}(\epsilon,v,-\lambda).
\end{align}
\end{subequations}
In the calculation of the gradients of such quantities, we notice that
\begin{subequations}
\begin{align}
 \left(\partial_\epsilon\varphi\right)(\epsilon,-v,\lambda) &=
 -\left(\partial_\epsilon\varphi\right)(\epsilon,v,-\lambda),\\
 \left(\partial_v\varphi\right)(\epsilon,-v,\lambda) &=
 \left(\partial_v\varphi\right)(\epsilon,v,-\lambda), 
\end{align}
\end{subequations}
for the response coefficients and
\begin{subequations}
\begin{align}
\left(\partial_\epsilon\braket{\hat{R}}^{(i)}\right)(\epsilon,-v,\lambda)&=
\left(\partial_\epsilon\braket{\hat{R}}^{(i)}\right)(\epsilon,v,-\lambda),\\
\left(\partial_v\braket{\hat{R}}^{(i)}\right)(\epsilon,-v,\lambda)&=
-\left(\partial_v\braket{\hat{R}}^{(i)}\right)(\epsilon,v,-\lambda), 
\end{align}
\end{subequations}
for the average charge and spin. The pseudo magnetic fields then write
as follows
\begin{subequations}
\begin{align}
\mathcal{B}_{I_L}(\epsilon,-v,\lambda) &=
 \mathcal{B}_{I_L}(\epsilon,v,-\lambda),\\
\mathcal{B}_{J_L}(\epsilon,-v,\lambda) &=
 \mathcal{B}_{J_L}(\epsilon,v,-\lambda).
\end{align}
\end{subequations}
These two relations can be directly checked by comparing the left and
right panels of Fig.~\ref{fig:pseudob1}. For $\lambda = 0$ (see
central panels) the above equations imply a symmetric shape of the
pseudo magnetic fields around the zero-bias axis.\\

\subsection{Gate voltage reversal}

We calculate now the pseudo magnetic fields for an inversion of the gate
voltage around $V_g = U/2$. In this case, we take as the new origin of
coordinates the point $(\alpha V_g,V_b) = (U/2,0)$ such that the factors
in Eq.~(\ref{eq:rates}) write
\begin{subequations}
\begin{align}
\gamma_r(\epsilon, v,\lambda) &= \sum_\sigma \tfrac{\Gamma_r(\lambda)}{2}
\left[ f(\epsilon_{r\sigma}-\tfrac{U}{2}) -
 f(\epsilon_{r\sigma}+\tfrac{U}{2})\right],\\ 
\beta_r(\epsilon, v,\lambda) &= \sum_\sigma \sigma
 \tfrac{\Gamma_r(\lambda)}{2} \left[ f(\epsilon_{r\sigma}-\tfrac{U}{2}) - 
 f(\epsilon_{r\sigma}+\tfrac{U}{2})\right],
\end{align}
\end{subequations}
where $\epsilon_{r\sigma} = \epsilon-\alpha_r v -\sigma B/2$. For these
new coordinates, the inversion of the gate voltage is then given by
$\epsilon \rightarrow -\epsilon$, and the above factors are transformed
according to 
\begin{subequations}
\begin{align}
\gamma_r(-\epsilon, v,\lambda) &= \gamma_{\bar{r}}(\epsilon, v,-\lambda),\\ 
\beta_r(-\epsilon, v,\lambda) &= -\beta_{\bar{r}}(\epsilon, v,-\lambda).
\end{align}
\end{subequations}
In consequence, the response coefficients obey the following relations  
\begin{subequations}
\begin{align}
 \varphi_n^{I_L}(-\epsilon,v,\lambda) &= -1 -
 \varphi_n^{I_L}(\epsilon,v,-\lambda),\\ 
 \varphi_{S_z}^{I_L}(-\epsilon,v,\lambda) &=
 \varphi_{S_z}^{I_L}(\epsilon,v,-\lambda),\\
 \varphi_n^{J_L}(-\epsilon,v,\lambda) &=
 \varphi_n^{J_L}(\epsilon,v,-\lambda),\\
 \varphi_{S_z}^{J_L}(-\epsilon,v,\lambda) &= -1 -
 \varphi_{S_z}^{J_L}(\epsilon,v,-\lambda). 
\end{align}
\end{subequations}
For the average charge and spin we obtain 
\begin{subequations}
\begin{align}
 \braket{\hat{n}}^{(i)}(-\epsilon,v,\lambda) &= 2
 -\braket{\hat{n}}^{(i)}(\epsilon,v,-\lambda),\\
 \braket{\hat{S}_z}^{(i)}(-\epsilon,v,\lambda) &= 
 \braket{\hat{S}_z}^{(i)}(\epsilon,v,-\lambda),
\end{align}
\end{subequations}
such that the pseudo magnetic fields present different symmetries with 
respect to a change in the gate voltage, i.e.
\begin{subequations}
\begin{align}
 \mathcal{B}_{I_L}(-\epsilon,v,\lambda) &=
 -\mathcal{B}_{I_L}(\epsilon,v,-\lambda),\\
 \mathcal{B}_{J_L}(-\epsilon,v,\lambda) &=
 \mathcal{B}_{J_L}(\epsilon,v,-\lambda).
\end{align}
\end{subequations}

\section{Maximum values of $\mathcal{B}_{I_L}$}
\label{app:roots}

In this section we calculate the point in which the pseudo magnetic
field is maximum, i.e. the position of the peak (1) in
Fig.~\ref{fig:field}. Our starting point is the explicit form of the
pseudo magnetic field at low bias. According to Eq.~(\ref{eq:bfield}),
the condition for an extremum point $(\chi_L,\chi_R)$ of the pseudo
magnetic field is the following:
\begin{subequations} 
\begin{align}
 \tanh \left(\frac{\chi_L}{2}\right) &=
 \frac{-3(1+\lambda)\partial_{\chi_L}f_L}
 {2+(1+\lambda)f_L+(1-\lambda)f_R},\\ 
 \tanh \left(\frac{\chi_R}{2}\right) &=
 \frac{-3(1-\lambda)\partial_{\chi_R}f_R} 
 {2+(1+\lambda)f_L+(1-\lambda)f_L},
\end{align}
\end{subequations}
where $f_r = f(\chi_r)$, $r = L,R$. Now we take the replacements $x =
1+e^{\chi_L}$ and $y = 1+e^{\chi_R}$, such that $x,y > 1$ and the above
equations read
\begin{subequations} 
\begin{align}
3(1+\lambda)y &= \frac{x-2}{x-1}(2xy+x+y+\lambda(y-x)),\\
3(1-\lambda)x &= \frac{y-2}{y-1}(2xy+x+y+\lambda(y-x)).
\end{align}
\end{subequations}
By solving the first equation, we obtain
\begin{equation}
y = -\frac{(1-\lambda)x(x-2)}{2x^2-2(3+\lambda)x+1+\lambda},
\label{eq:yfactor}
\end{equation}
and plugging this into the second equation we obtain the following
quartic equation for $x$:
\begin{equation}
ax^4+bx^3+cx^2+dx+e = 0,
\end{equation}
with
\begin{subequations}
\begin{align}
a &= 2\lambda-6,   & b &= -\lambda^2-4\lambda+29,\\
c &= 2\lambda^2-34,& d &= -2(\lambda+1)^2,\\
e &= (\lambda+1)^2.
\end{align}
\end{subequations}
For the particular case of symmetric junctions, the solutions compatible
to the condition $x>1$ are $x = (5+\sqrt{33})/4$ and $x = (7+\sqrt{37})/6$
respectively. Now if we use these values in Eq.~(\ref{eq:yfactor}), the
only solution that fulfills the condition $y>1$ is $y = x =
(5+\sqrt{33})/4$. Therefore, for symmetric junctions, the maximum of the
pseudo magnetic field is located at the point $\chi_L = \chi_R =
\ln[(1+\sqrt{33})/4]$.


\begin{thebibliography}{99}
\bibitem{note1}	The instantaneous component simply consists of the
	time-averaged stationary flow and is not discussed here.
\bibitem{note2} Since the surface integral would demand a precise
	knowledge of the pseudo magnetic field over the whole enclosed
	area, we calculate the asymptotic pumped charge as the contour
	integral of the vector potential introduced in
	Eq.~(\ref{eq:pseudoa}). This has the advantage that we can first
	take the limit $\delta\chi \gg 1$ (while keeping $\delta\chi \ll
	U/T$) in the definition of the vector potential and then
	calculate the pumped charge.
\bibitem{note3} The external magnetic field is indicated by the regular
	font $B$ and should not be confused with the pseudo magnetic
	field $\mathcal{B}$.
\end{thebibliography}
\end{document}